\documentclass[3p]{elsarticle}

\usepackage{amssymb}
\usepackage{amsmath}

\usepackage{hyperref}
\usepackage{booktabs}
\usepackage{array}
\usepackage{longtable}

\usepackage{CJKutf8}
\newcommand{\zh}[1]{{\CJKfamily{gbsn} #1}}
\newcommand{\jp}[1]{{\CJKfamily{min} #1}}

\usepackage{algorithm}

\usepackage{algorithmicx}
\usepackage{algpseudocode}
\usepackage{multirow}

\usepackage{graphicx}
\usepackage{subcaption}
\usepackage{lineno}

\usepackage{makecell}
\usepackage{hhline}
\usepackage{adjustbox}

\makeatletter
\renewcommand{\ALG@beginalgorithmic}{\footnotesize}
\algrenewcommand{\algorithmiccomment}[1]{// #1}
\makeatother

\makeatletter
\def\ps@pprintTitle{%
  \let\@oddhead\@empty
  \let\@evenhead\@empty
  \let\@oddfoot\@empty
  \let\@evenfoot\@empty}
\makeatother

\journal{}

\begin{document}

\begin{frontmatter}

\title{Predictive Modeling: BIM Command Recommendation Based on Large-Scale Usage Logs}

\author[inst1,inst2]{Changyu Du} 
\ead{changyu.du@tum.de}

\affiliation[inst1]{
            organization={Chair of Computing in Civil and Building Engineering, Technical University of Munich},
            country={Germany}}
\affiliation[inst2]{
            organization={TUM Georg Nemetschek Institute},
            country={Germany}}

\author[inst1,inst2]{Zihan Deng}
\author[inst1,inst2]{Stavros Nousias}
\author[inst1,inst2]{André Borrmann}

\begin{abstract}
The adoption of Building Information Modeling (BIM) and model-based design within the Architecture, Engineering, and Construction (AEC) industry has been hindered by the perception that using BIM authoring tools demands more effort than conventional 2D drafting. To enhance design efficiency, this paper proposes a BIM command recommendation framework that predicts the optimal next actions in real-time based on users' historical interactions. We propose a comprehensive filtering and enhancement method for large-scale raw BIM log data and introduce a novel command recommendation model. Our model builds upon the state-of-the-art Transformer backbones originally developed for large language models (LLMs), incorporating a custom feature fusion module, dedicated loss function, and targeted learning strategy. In a case study, the proposed method is applied to over 32 billion rows of real-world log data collected globally from the BIM authoring software Vectorworks. Experimental results demonstrate that our method can learn universal and generalizable modeling patterns from anonymous user interaction sequences across different countries, disciplines, and projects. When generating recommendations for the next command, our approach achieves a Recall@10 of approximately 84\%. The code is available at: \url{https://github.com/dcy0577/BIM-Command-Recommendation.git}
\end{abstract}



\begin{keyword}
Building Information Modeling (BIM) \sep Sequential recommendation \sep BIM logs processing \sep Transformer \sep Large Language Model (LLM)\sep Machine learning


\end{keyword}

\end{frontmatter}


\section{Introduction}

Modern BIM authoring tools integrate various disciplines and systems within facility design, enabling cohesive and collaborative workflows. However, this integration comes at the cost of increased complexity of the user interface and proliferation of authoring commands, which prolong modeling times and increase the likelihood of user errors \cite{Du:2024:command_recommender}. Despite ongoing efforts by software vendors to simplify user interfaces, designers often encounter steep learning curves, relying heavily on trial-and-error methods to locate and apply suitable commands for their tasks. Even experienced professionals may struggle to translate their expertise into the intricate and multifaceted command flows required during the BIM authoring process. Therefore, a predictive decision-support system capable of recommending optimal next actions within the BIM authoring tools could significantly enhance efficiency, reduce modeling time, and minimize the risk of errors.

Related research on command prediction based on BIM logs \cite{Du:2024:command_recommender,GAO2022104026,pan2020_log_mining,doi:10.1061/9780784482865.033} aims to model design patterns by mining sequential records of events collected during the use of BIM authoring software \cite{JANG2023102079}. Existing studies often rely on custom or smaller-scale BIM log datasets, typically predicting single-step commands limited to specific scopes or design stages. Moreover, the employed algorithmic approaches are based on basic sequence prediction models, without exploring more advanced model architectures that could fuse and leverage additional information available in BIM logs to enhance command prediction.

\begin{figure}[!ht]
    \centering
    \includegraphics[width=0.7\linewidth]{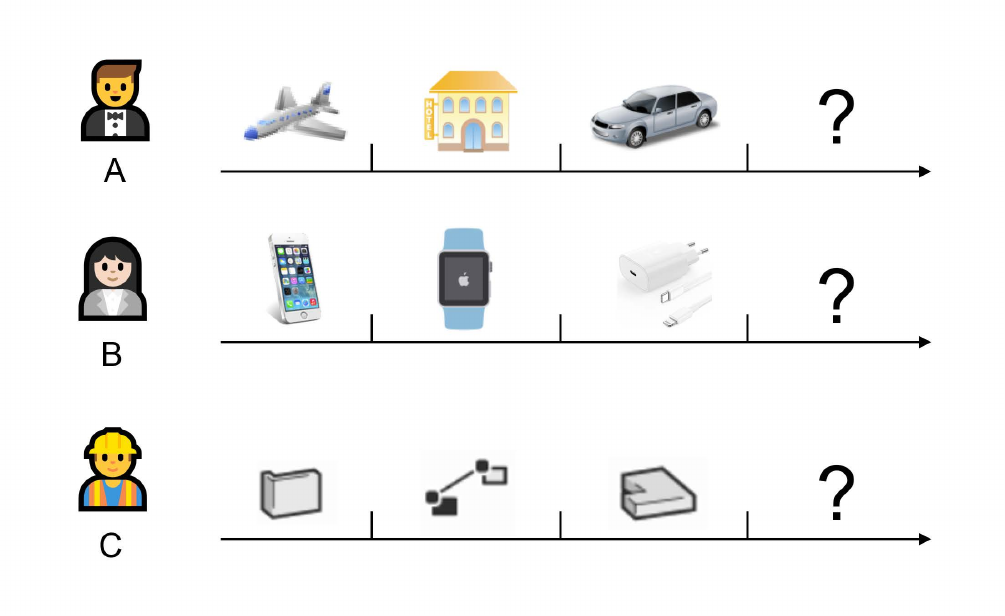}
    \caption{Examples of two sequential recommendation problems in e-commerce scenario \cite{ijcai2019p883} and the similarities to BIM command recommendation: (a) After A books a flight, a hotel, and rents a car, what will his next actions be? (b) After B purchases an iPhone, an Apple Watch, and a charging cable, what will she buy next? (c) After C draws a wall, moves it, and creates a floor slab in the BIM authoring tool, what will he do next?}
    \label{fig:example_srs}
\end{figure}

The deep sequential recommendation system offers a promising approach to improve BIM command prediction. These systems have been extensively studied and applied in domains such as e-commerce and social networks \cite{wang2023multitaskdeeprecommendersystems}, leveraging advanced Transformer-based models to learn complex behavioral patterns from massive user-item interaction histories and rich contextual information (e.g., item price, timestamps, categories, etc.). Such systems provide personalized recommendations that help users efficiently filter through massive catalogs to find the most relevant items and predict subsequent actions \cite{ijcai2019p883}. We are addressing similar challenges faced during the BIM authoring process, as illustrated in Figure~\ref{fig:example_srs}.

Inspired by such systems, this paper proposes a BIM command recommendation framework to address the limitations of previous studies. By treating commands as recommendable items, the framework aims to generate real-time recommendations throughout the whole software usage life cycle. The research questions explored in this study can be summarized as follows:

\begin{itemize}
    \item Data processing and Enhancement: How can large-scale raw log data be effectively filtered and enriched to provide high-quality input suitable for command recommendation models?
    \item Model Architecture Design: How to design an effective Transformer-based BIM command recommendation model that can learn universal and generalizable modeling patterns from designers worldwide?
    \item Model Evaluation: How can the performance and effectiveness of the designed model architecture be evaluated?
    \item Deployment and Integration: How can the model be deployed and integrated into the BIM authoring process to provide real-time predictive decision support?
\end{itemize}

This paper addresses these questions by proposing a BIM command recommendation framework. We develop a comprehensive filtering and enhancement method for large-scale raw BIM log data, addressing critical challenges not covered by previous studies. 

A novel command recommendation model is proposed, which builds upon the state-of-the-art Transformer backbones originally developed for large language models (LLMs) and incorporates custom feature fusion modules, dedicated loss functions, and targeted learning strategies. Thanks to the advanced data augmentation method, the trained model can not only recommend individual commands but also potential workflows that package multiple consecutive action steps.

In a comprehensive case study, the proposed method is applied to over 32 billion rows of real-world log data collected globally from the BIM authoring software Vectorworks, surpassing the scale of all previous studies. Extensive evaluations and experiments are conducted on the proposed method to analyze its effectiveness and limitations. Furthermore, a software prototype is implemented to deploy and integrate the trained model, enabling real-time command recommendations during the BIM authoring process. 

The proposed end-to-end framework holds practical significance in guiding the implementation of predictive modeling within engineering software.

\section{Background and Related work}

\subsection{Command prediction based on BIM logs}

BIM logs are a chronological record of events automatically generated during the use of BIM authoring software. The techniques and methods used to analyze these logs can be collectively referred to as BIM log mining \cite{JANG2023102079}. This includes but is not limited to examining designers' social networks to understand collaboration patterns \cite{PAN2022101758,ZHANG201831}, identifying the productivity of modelers \cite{YARMOHAMMADI201717,PAN2020102997,asce_log_mining}, analyzing the creativity or quality of designs \cite{GAO2023104804,NI2025105992}, and enhancing the reproducibility of the modeling process \cite{jang2021logging}, etc. One important use case is leveraging the sequential user operations recorded in the logs to predict the next command. 

Pan et al. extracted command sequences from Revit log files and grouped them into 14 classes that attempted to summarize the generic intent of the different commands at a high level \cite{pan2020_log_mining}. Eventually, a long-short-term memory network (LSTM) \cite{LSTM} was employed to predict the class labels of upcoming commands. However, their approach is restricted to predicting limited command categories rather than individual commands. \cite{radnia2021sequence} used a similar concept to predict command classes based on Revit log files. They built models at different scales using different numbers of LSTM layers and compared the prediction results. Guo et al. developed a custom log in Rhino to combine modeling commands with their resulting 3D models, proposing the command-object graph to represent the modeling design behavior \cite{GAO2021103967}. Their subsequent research \cite{GAO2022104026} extracted paths from the graphs to compose extensive command sequences, and used the native Transformer model \cite{transformer} to achieve the instance-level command prediction. Although the custom log can extract command sequences more effectively by integrating specific information of model elements compared to native logs, its limitations lie in limited public access and the necessity for manual updates \cite{JANG2023102079}. This results in a constrained dataset that may not accurately reflect real-world software usage. Furthermore, the basic Transformer model used in the study does not fully leverage the additional meta-information in the log files.

The CommunityCommands \cite{10.1145/1970378.1970380,CommunityCommands} for AutoCAD provides personalized command recommendations using an item-based collaborative filtering algorithm. It evaluates the importance of commands through a command frequency-inverse user frequency (cf-iuf) rating \cite{CommunityCommands}, combining personal usage frequency with community-wide rarity. Recommendations are based on the cosine similarity between unused commands and those already used by the active user. Despite its innovative statistical approach, an evaluation of data from 4,033 users revealed limited predictive accuracy, with a 21\% of hit rate among the top 10 recommendations.

\subsection{Transformers for sequential recommendation}

Recent advancements in large language models (LLMs) have demonstrated remarkable capabilities in modeling sequential data through the Transformer architecture. Originally developed for Natural Language Processing (NLP), Transformers have been successfully adapted to recommendation systems due to their ability to capture complex dependencies in sequential user interactions \cite{10.1145/3357384.3357895,8594844,t4rec}. The attention mechanism enables effective modeling of user behavior dynamics, making it well-suited for scenarios where user actions and interests evolve across sessions \cite{Du:2024:command_recommender}.

While NLP Transformers typically process text through tokenizers and task-specific prediction heads, recommendation systems adapt this architecture by incorporating meta-information such as user comments, product descriptions, and prices in e-commerce domains \cite{10.1145/3523227.3546777}. Our study leverages the core Transformer architecture to model sequential dependencies among commands, replacing NLP-specific components with a custom feature fusion module and a prediction head tailored for command recommendation.

Our implementation builds on Transformer4Rec \cite{t4rec}, an open-source framework that bridges NLP and sequential recommendation systems using Transformer backbones from models like XLNet \cite{xlnet}. This framework has demonstrated state-of-the-art performance in news and e-commerce domains. However, our research explores BIM command recommendation, which presents unique challenges including limited meta-information, varying session lengths, and long-tail command distributions.

\subsection{Summary and research gaps}
\label{gaps}

In summary, we identify the existing research gaps in the literature as follows:
\begin{itemize}

    \item From the algorithm perspective, 
    existing research primarily employs statistical methods or basic sequence prediction models, overlooking advanced deep sequential recommendation systems that have shown success in other domains. Moreover, current studies focus on predicting single-step command instances or classes. However, design tasks typically involve multi-command workflows, making it more practical to enhance prediction granularity and recommend optimal workflows to users.
    \item From the system integration perspective, current studies have not proposed an end-to-end pipeline that seamlessly integrates the prediction model into BIM authoring software for real-time command recommendation.
    \item From the data perspective, existing studies often rely on small-scale datasets generated through bespoke logging mechanisms, neglecting the engineering challenges of processing real-world log data at the billions-scale in its raw format. This limitation also constrains the scalability of existing data enhancement methods based on customized loggers.
    
\end{itemize}

\newpage
\section{Methodology}
\label{sec1}

\begin{figure}[ht!]
    \centering
    \includegraphics[width=1\linewidth]{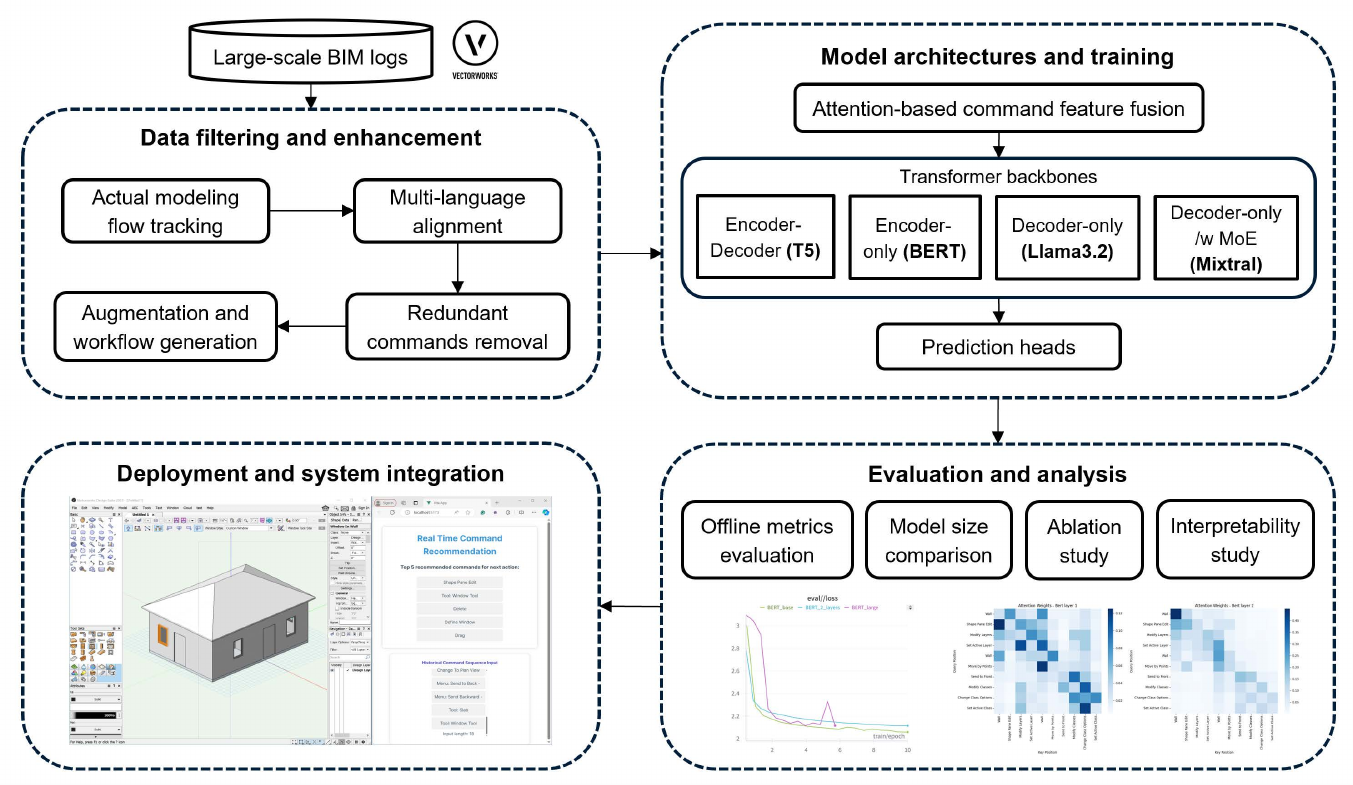}
    \caption{The overall research framework}
    \label{fig:framework}
\end{figure}

To bridge the gaps outlined in Section \ref{gaps}, this study proposed a framework for large-scale BIM command recommendation, as illustrated in Figure \ref{fig:framework}.

We first propose a comprehensive data processing approach to address issues such as information redundancy, multilingual data, and mismatch errors commonly encountered in large-scale real-world BIM logs, enabling the capture of users' true software usage operations. Furthermore, our data augmentation method leverages LLMs combined with software documentation knowledge to supplement additional command meta-information. Notably, this approach does not rely on customized loggers and can be directly applied to standardized raw logs. Inspired by the BPE (Byte Pair Encoding) algorithm \cite{BPE}, we aggregate frequently occurring consecutive commands to form diverse workflows, expanding the recommendation scope. This enables the model to predict not only individual commands but also potential multi-step workflows.

The proposed model architecture effectively aggregates the augmented command meta-information and explores incorporating state-of-the-art LLM architectures into the command recommendation context. Through a targeted learning strategy and special loss functions, the model's performance is enhanced. Extensive experiments are conducted to analyze model performances. A software architecture is also introduced to deploy the trained model along with the data processing pipeline into the BIM authoring scenario, achieving end-to-end real-time command recommendations.

\subsection{Data filtering and enhancement}
\label{datafilteringworkflow}

\begin{figure}[ht!]
    \centering
    \includegraphics[width=\linewidth, trim=8mm 197mm 53mm 11mm, clip]{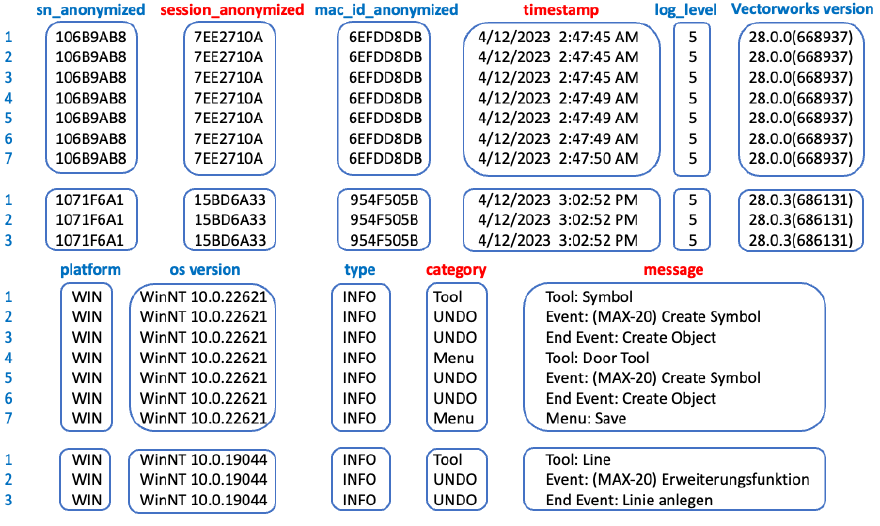}
    \caption{Overview of the Vectorworks native log file (anonymized). Blue fields represent system-relevant data, while red fields denote design-critical information, including the anonymized session ID indicating an independent modeling session (\texttt{session\_anonymized}), the timestamp marking the log entry (\texttt{timestamp}), the predefined category associated with the command (\texttt{category}), and the command prefix along with name (\texttt{message}).}    
    \label{fig:native_logs}
\end{figure}

Previous studies have primarily focused on developing data processing pipelines based on Revit logs; a comprehensive review is provided in \cite{JANG2023102079}. However, these studies have not addressed more challenging issues related to logs from other BIM authoring software, which often exhibit multilingual content, a mix of high- and low-level commands, and excessive complexity. As an example, Figure \ref{fig:native_logs} illustrates a snippet of the Vectorworks log containing two sessions in different languages.

Additionally, existing data processing methods, constrained by the scale of data they handle, fail to account for biases introduced during large-scale logging and global software distribution. This includes missing log entries, corrupted text, multiple languages, misaligned command IDs, etc. Additionally, existing data augmentation methods often require bespoke loggers or manual efforts to enrich the log contents, lacking automated solutions that can be directly applied to standard native BIM logs. 

\begin{figure}[ht!]
    \centering
    \includegraphics[width=1\linewidth, trim=7mm 155mm 7mm 10mm, clip]{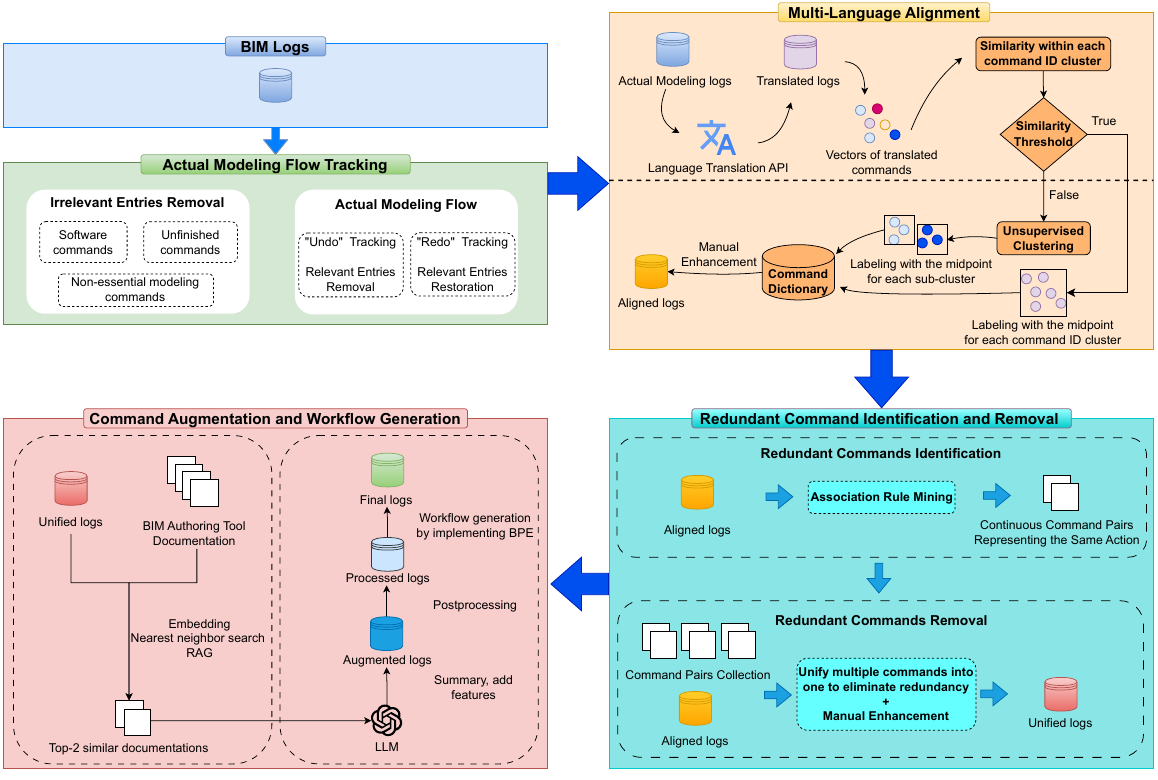}
    \caption{Multi-stage data filtering and enhancement method, including Actual Modeling Flow Tracking (Section \ref{actualmodelingflowtracking}), Multi-language Alignment (Section \ref{multilanguagealignment}), Redundant Command Identification and Removal (Section \ref{redundantcommand}), and Command Augmentation and Workflow Generation (Section \ref{augmentationandpostprocessing})}
    \label{fig:workflow}
\end{figure}

To overcome these challenges, we proposed a comprehensive multi-stage data filtering and enhancement pipeline (illustrated in Fig. \ref{fig:workflow}), consisting of four modules:

\begin{itemize} 
\item \texttt{Actual Modeling Flow Tracking}: Filters out irrelevant entries and tracks redo/undo operations to remove or restore relevant entries. 
\item \texttt{Multi-Language Alignment}: Resolves challenges of multilingual content and misaligned command IDs by standardizing them into a unified English representation.
\item \texttt{Redundant Command Identification and Removal}: Remove redundant log entries to ensure each action is uniquely represented by a single log entry.
\item \texttt{Command Augmentation and Workflow Generation}: Enrich command meta-information by leveraging LLMs to summarize domain knowledge from software documentation. BPE is then applied to further generate multi-command workflows.
\end{itemize}

The concrete implementation of the data filtering and enhancement method in a case study is illustrated in Section \ref{data_processing} along with example outcomes.

\subsubsection{Actual modeling flow tracking}
\label{actualmodelingflowtracking}

The raw BIM logs require cleaning and preprocessing to accurately trace the actual modeling flow. The initial step focuses on removing irrelevant entries from both semantic and statistical perspectives. Entries related to internal software events, such as error alerts that do not involve user actions were removed. Additionally, commands that frequently occur but have minimal significance for BIM authoring and recommendations, such as zooming and panning, were excluded. Lastly, entries corresponding to aborted or unfinished commands, as well as those with extremely low occurrence rates, were filtered out to ensure the dataset's relevance and clarity.

Another aspect of the process focuses on handling \textit{Undo} and \textit{Redo} logic. Undoing or redoing a command generates several new entries in the log, rather than removing or restoring the existing entry. This results in redundant log entries that do not accurately represent the actual workflow. An example can be found in the native log in Figure \ref{log_compare}.
To address this, we developed an algorithm that ensures that all undo- or redo-related commands are accurately identified and removed from the log, preserving the true sequence of the user's operations.
As illustrated in Alg.~\ref{algorithm_undo_redo}, this algorithm implements the matching and removal of “Redo” and “Undo” commands by maintaining two command lists (\textit{recent\_commands} and \textit{recent\_undone\_commands}) and a set of indices to be removed (\textit{to\_remove\_indices}). It traverses log entries sequentially in chronological order within each session. Whenever a normal command appears, its name and corresponding index are recorded in \textit{recent\_commands}. When an “Undo” command is detected, the algorithm searches backward through \textit{recent\_commands} to find the relevant command it applied on; upon a match, both their indices are added to \textit{to\_remove\_indices}, and the normal command is moved to \textit{recent\_undone\_commands}. If a “Redo” command is encountered, the algorithm first searches backward in \textit{recent\_undone\_commands} for a match. If found, that command is restored from the to-be-removed state, and the “Redo” command itself is marked for removal. If no match is found, only the “Redo” command is marked for removal. After all log entries are traversed, any commands marked for removal are deleted.

\subsubsection{Multi-language alignment}
\label{multilanguagealignment}

The global usage of BIM authoring software results in multilingual logs, which cannot be directly used for model training due to inconsistent representations of the same command across languages. Despite each command being associated with an ID, in real-world scenarios, inconsistencies arise due to variations in country-specific usage, regional settings, and different versions of software. For instance, the same command name may correspond to different IDs, and conversely, the same ID might represent different command names, as demonstrated in Table~\ref{tab:language_dic}. These inconsistencies, stemming from diversified global customization, complicate the processing and analysis of BIM logs. As a result, relying on command IDs for representing unique commands is neither robust nor accurate. 

To address these challenges, we propose a novel approach that aligns the representation of commands across different languages by assigning a standardized English name to equivalent commands. The English names eventually replace unreliable IDs as the unique identifier for each command.

Specifically, a language translation API is initially employed to translate commands into English. However, inconsistencies persist, such as variations in letter case and synonymous translations of the same command. To resolve these issues, a text embedding model is used to convert the translated commands into vector embeddings. For commands sharing the same ID, we calculate their pair-wise average cosine similarity and set a threshold. If the similarity exceeds this threshold, this group of commands is deemed equivalent, as the names associated with this ID are semantically very similar to form a distinct cluster in high-dimensional space. Therefore, the translated English name corresponding to the midpoint vector is assigned to the ID. 

If the similarity falls below the threshold, we assume that the ID is shared by multiple different commands, as the English names obtained by the translation API show significant semantic differences. In such cases, the unsupervised clustering algorithm DBSCAN \cite{DBSCAN1, DBSCAN2} is employed to further group the commands into sub-clusters. Each sub-cluster is subsequently labeled using the command name of its midpoint vector. The process is outlined in Alg. \ref{algorithm_language}. The example outcomes from this module are illustrated in Table \ref{tab:language_dic}.

\subsubsection{Redundant command identification and removal}
\label{redundantcommand}

In BIM logs, a user operation often consists of multiple lo\textit{}g entries that collectively correspond to the same action. For instance, as illustrated in Figure \ref{fig:native_logs}, when an operation "Symbol" or "Door Tool" is performed, a high-level command categorized as Tool is generated to record the selection of the corresponding UI button (\textit{Tool: Symbol} or \textit{Tool: Door Tool}). Simultaneously, the start and completion of the operation are recorded separately by the triggered low-level commands (\textit{Event: Create Symbol} and \textit{End Event: Create Object}).

To improve the efficiency and accuracy of the recommendation system while reducing log redundancy, it is essential to retain only one command to represent each user operation, ideally the high-level command accessible through the UI. However, no official documentation exists to define the relationship between commands that represent the same operation (e.g., which high-level command is triggering which low-level command). Additionally, triggering relationships are highly dynamic and may follow one-to-one, one-to-many, or many-to-many patterns. For instance, Figure \ref{fig:native_logs} shows that the same low-level commands can be triggered by different high-level commands \textit{Tool: Symbol} and \textit{Tool: Door Tool}. 
Since multiple command log entries for the same operation typically occur consecutively, we employ an advanced statistical method, Association Rule Mining (ARM) \cite{arm}, to establish linkages between them.

Association Rule Mining is a data mining technique used to identify patterns and relationships among items within large datasets. This technique can be expressed mathematically as:
\begin{equation}
\text{\textit{Support}}(X \cap Y) = \frac{|T(X \cap Y)|}{|D|}
\end{equation}

\begin{equation}
\text{\textit{Confidence}}(X \rightarrow Y) = \frac{\text{\textit{Support}}(X \cap Y)}{\text{\textit{Support}}(X)}
\end{equation}

where \(|T(X \cap Y)|\) is the count of command pairs \(X\) and \(Y\) that frequently occur together, \(|D|\) is the total number of commands in the dataset, and \(\text{\textit{Support}}\) and \(\text{\textit{Confidence}}\) measure the strength and reliability of the associations.

We begin by calculating the \textit{Confidence} for the association between each high-level command and its subsequent low-level commands based on the timestamps in the dataset. A confidence threshold is established to determine significant relationships between commands. If the confidence value falls below the threshold, the commands are considered to represent different operations. For commands with confidence values exceeding the threshold, further refinement is conducted by identifying the top most likely subsequent low-level commands. Since different modes of high-level commands in the BIM authoring tool may trigger varying sets of subsequent low-level commands, additional testing is performed by manually setting different operational modes and observing the resulting commands. This process helps establish a mapping, which is a collection of command pairs, each consisting of a high-level command and its corresponding low-level commands. The process is summarized in Alg. \ref{mapping}.

Based on the obtained mapping, we iterate through the aligned logs from the previous step to keep the recommendable high-level commands
and remove the low-level commands triggered by them. Additionally, we remove high-level commands that do not have corresponding low-level command records, as this indicates that the high-level command was not fully executed or completed. The example outcomes from this module are presented in Figure \ref{log_compare}.

\subsubsection{Command augmentation and workflow generation}
\label{augmentationandpostprocessing}

\begin{figure}
    \centering
    \includegraphics[width=1\linewidth]{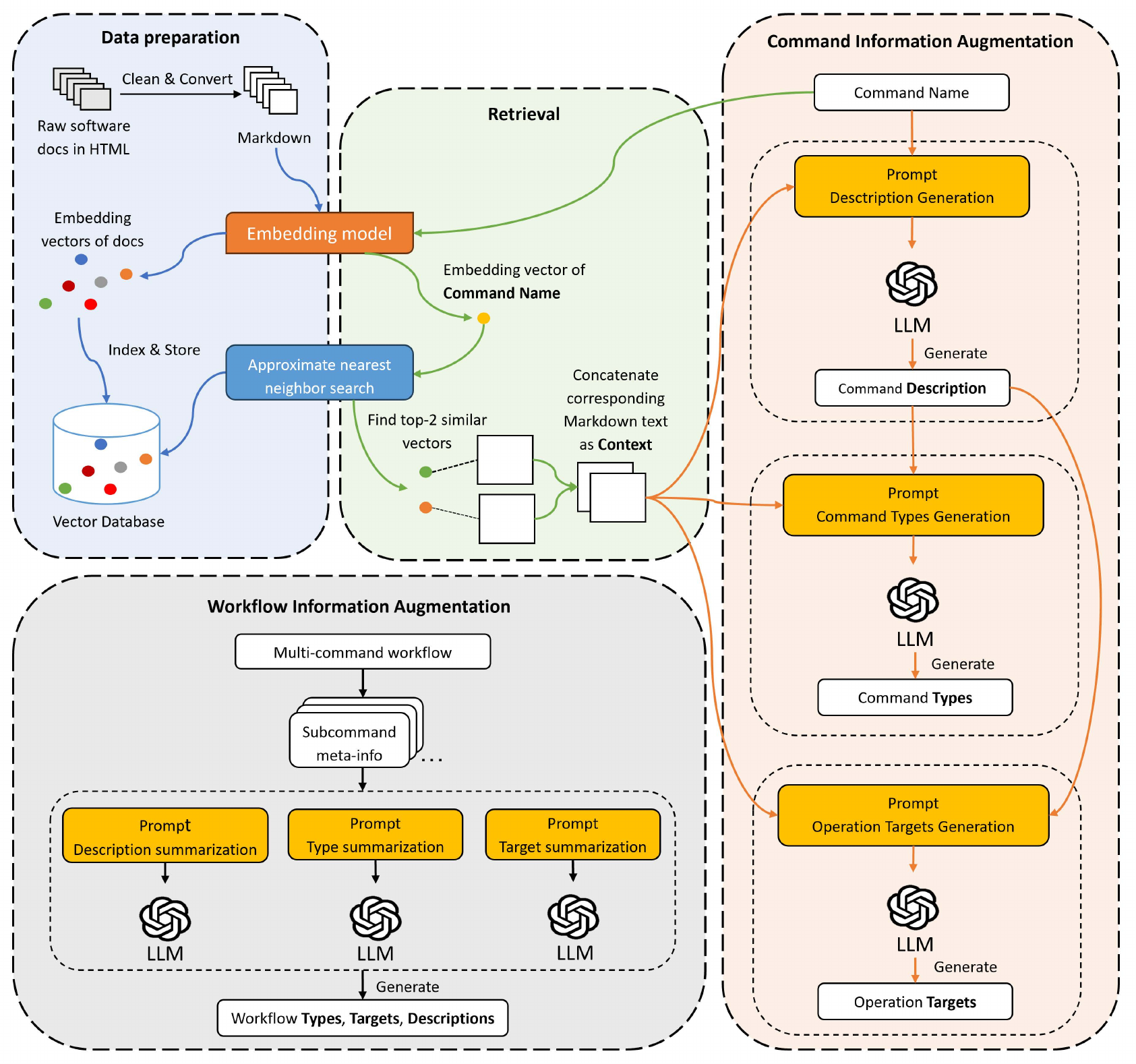}
    \caption{Retrieval-Augmented Generation (RAG) workflow extended from \cite{Du_Copilot_24} for log information augmentation. }
    \label{fig:information_augmentation}
\end{figure}

Commands in BIM logs are often recorded in short and ambiguous names, lacking rich meta-information compared to the items in e-commerce scenarios. However, the software documentation provided by BIM authoring tools offers detailed descriptions for each command, presenting an opportunity to enrich the information in raw logs. Inspired by \cite{Du_Copilot_24}, we developed a custom Retrieval-Augmented Generation (RAG) \cite{RAG} workflow that enriches log information by combining world knowledge from Large Language Models (LLMs) and domain-specific knowledge from documentation, which is illustrated as Figure \ref{fig:information_augmentation}. Unlike existing data augmentation methods, which require developing custom loggers, our approach can be directly applied to standard native BIM logs, providing a scalable and efficient solution.

The process begins by collecting and cleaning HTML files from the BIM authoring tool's online documentation, which are then converted into Markdown format. This documentation is subsequently processed using a text embedding model to generate vector representations of its contents. These vectors are stored in a vector database for efficient retrieval \cite{Du_Copilot_24}. Simultaneously, command names are transformed into vector representations using the same embedding model. By calculating the semantic similarity between these vectors, the two most relevant documentation contents are retrieved for each command name. The retrieved content is automatically aggregated and summarized by the LLM to form the command descriptions. 

Previous research \cite{pan2020_log_mining} manually classified commands from Revit journal files into 14 distinct classes. Inspired by their approach, commands are further classified by the LLM into different types based on the summarized description. Additionally, their operation targets, representing the specific objects affected by the commands, are inferred by the LLM to further augment command meta-information.

In our recommender system, we would like to recommend not only individual commands but also potential workflows that package multiple consecutive action steps.
Our approach draws inspiration from subword tokenization techniques Byte Pair Encoding (BPE) \cite{BPE} from NLP, which iteratively merges frequently occurring consecutive character pairs into single units to form meaningful tokens. By treating commands in logs as "words" in natural language, we apply similar logic to generate multi-command workflows by capturing combinations of consecutive commands representing frequently
used action sequences. However, it has to be emphasized that commonly available trained LLMs, which are trained on large corpus of human-language text, cannot be applied for our purposes, as the difference between human language and a sequence of commands is simply too big.

Specifically, we first count all commands and their frequencies to generate an initial command vocabulary. Then, we compute the frequencies of adjacent command pairs and merge the most frequent pair to update the vocabulary. This merging process is iteratively repeated until the predefined vocabulary size is reached. This aggregation incorporates multi-command workflows into the command vocabulary, enabling the downstream recommendation model trained on this vocabulary to expand its prediction scope to users' multi-step operations. Additionally, as illustrated in Figure \ref{fig:information_augmentation}, the aggregated workflows undergo the augmentation process as individual commands. For each command that constitutes a workflow, the meta-information is extracted, aggregated based on the sequence of commands within the workflow, and summarized by the LLM to generate context-aware meta-information, including descriptions, types, and targets for workflows. Table \ref{merged_table} demonstrates example commands and workflows along with their LLM-augmented meta-information.

\subsection{Model architecture}

\begin{figure}[ht]
    \centering
        \includegraphics[width=1\linewidth]{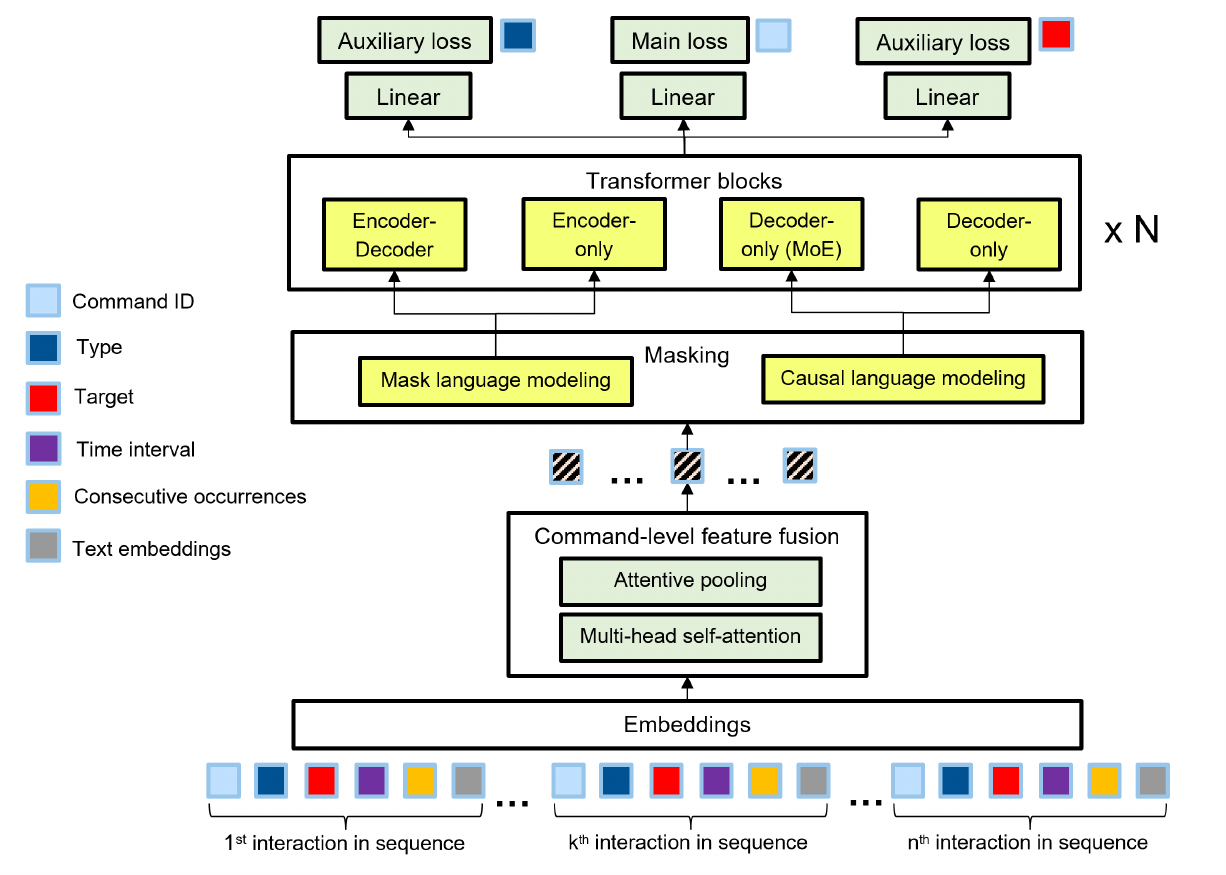}
    \caption{The overall model architecture in this study. Green blocks indicate that they appear simultaneously within their respective modules. The yellow parallel blocks indicate that only one can appear within the given module, representing several alternatives for the module's configuration.}
    \label{model_architecture}
\end{figure}

This section introduces the overall structure of the model, as illustrated in Figure \ref{model_architecture}, which is inspired by Transformer4Rec \cite{t4rec}. Given an input sequence, our model initially employs embedding layers to project each command-level meta-information into a shared-dimensional embedding space. For interaction at each time step, these different features are fused using the attention mechanism. The resulting updated sequence is masked to exclude the commands requiring prediction and fed into stacked Transformer blocks to learn the contextual and sequential dependencies. We compare four types of Transformer architectures widely used in state-of-the-art large language models (LLMs): encoder-only, decoder-only, decoder-only with MoE, and encoder-decoder. The output sequence embeddings are passed through three parallel linear branches to perform multi-task, multi-class predictions, with the main objective of recommending the next command instance IDs.

Compared to the Transformer4Rec baseline presented in \cite{Du_Copilot_24}, our model incorporates the following enhancements: (1) Replacing the original backbones with the state-of-the-art Transformer blocks from LLMs. (2) Employing attention mechanisms to fuse command-level features. (3) Adopting a multi-task learning approach to improve the accuracy of the primary task (next command instances prediction). (4) Mitigating the issue of label imbalance by using the focal loss \cite{focal_loss}.
Detailed explanations of each module are provided in the subsequent sections.

\subsubsection{Mathematical description}
\label{problem_statement}
Let $\mathcal{C} = \{c_1, c_2, \ldots, c_{|\mathcal{C}|}\}$ be a set of commands available in the BIM authoring software. 
Each command $c \in \mathcal{C}$ is associated with a set of \emph{atomic features} (or meta-information), such as command type, command target, timestamps, text descriptions, or other relevant attributes extracted from BIM logs. We denote these features collectively as $\mathbf{X}_c$. 
Consider a modeling session $S = \bigl[c_{1}^{(s)},\, c_{2}^{(s)},\, \ldots,\, c_{t}^{(s)},\, \ldots,\, c_{n}^{(s)}\bigr]$,
which represents a sequence of command interactions in chronological order during the session. Here, $c_{t}^{(s)} \in \mathcal{C}$ denotes the command interacted with at time step $t$ in session $S$, and $n$ is the length of the session. 

Given the $S$, the goal is to predict the next command 
\(
c_{n+1}^{(s)}
\)
that will be interacted with at time step $n+1$. Formally, this can be viewed as modeling the probability distribution over all possible commands at the next step, conditioned on the session history $S$ where each command is enriched with the aggregated meta-information:
\begin{equation}
p\bigl(c_{n+1}^{(s)} = c \mid S), \quad \forall\, c \in \mathcal{C}.
\end{equation}

In other words, sequential recommendation seeks to estimate the likelihood of each command being the next interaction within the current modeling session. By incorporating the additional meta-information into the sequence modeling process, we can enrich the representation of the session context and improve the accuracy of next-command prediction.

\subsubsection{Attention-based feature fusion}
\label{sec:att_fusion}

In the input sequence, each command is not only represented by a unique command ID, but also possesses features derived from various meta-information obtained by the data processing and augmentation pipeline, including categorical features (e.g., command types, categories, targets), continuous features (e.g., time interval, consecutive occurrences), and pretrained semantic embedding
acquired by inputting the command description into the state-of-the-art text-embedding-3-large model \cite{openai2024embedding}.
Formally, a command $c_t^{(s)}$ in the modeling session $S$ at time step $t$ is associated with a collection of $K$ distinct meta-information features:
\begin{equation}
\mathbf{X}_{c_t^{(s)}} \;=\;
\bigl\{
  \mathbf{x}_{c_t^{(s)}}^{(1)},\,
  \mathbf{x}_{c_t^{(s)}}^{(2)},\,
  \ldots,\,
  \mathbf{x}_{c_t^{(s)}}^{(K)}
\bigr\}
\end{equation}

To enrich the representation of each command \(c_{t}^{(s)}\), we propose an attention-based feature fusion module that effectively aggregates these meta-information features. 

We first map each feature in $\mathbf{X}_{c_t^{(s)}}$ to a common latent space using separate learnable linear projections so that all features become dimensionally consistent as $D$, resulting in a new collection of projected features at time step $t$: 
$
\tilde{\mathbf{X}}_{c_t^{(s)}} =
\bigl\{
  \tilde{\mathbf{x}}_{c_t^{(s)}}^{(1)},\,
  \ldots,\,
  \tilde{\mathbf{x}}_{c_t^{(s)}}^{(K)}
\bigr\},
$
Next, we treat the $\tilde{\mathbf{X}}_{c_t^{(s)}} \in \mathbb{R}^{K \times D}$ as query, key, and value and apply the Multi-head Self-Attention \cite{transformer} to fuse the features within the collection. The fundamentals of the attention mechanism are explained in Section \ref{transformer_backbone}.
\begin{equation}
\label{eq:multihead}
\widehat{\mathbf{X}}_{c_t^{(s)}}
\;=\;
\mathrm{MultiHeadAttention}\Bigl(
  \tilde{\mathbf{X}}_{c_t^{(s)}},
  \,\tilde{\mathbf{X}}_{c_t^{(s)}},
  \,\tilde{\mathbf{X}}_{c_t^{(s)}}
\Bigr)
\end{equation}
This fusion step leverages the self-attention mechanism to model pairwise interactions among the different feature types in 
\(\tilde{\mathbf{X}}_{c_t^{(s)}}\). Specifically, the multi-head attention module learns weights indicating the relative importance of each feature with respect to the others. Features that are highly relevant to one another are assigned higher weights, allowing them to contribute more strongly to each other's updated representations. Consequently, the output 
$
\widehat{\mathbf{X}}_{c_t^{(s)}} =
\bigl\{
  \widehat{\mathbf{x}}_{c_t^{(s)}}^{(1)},\,
  \ldots,\,
  \widehat{\mathbf{x}}_{c_t^{(s)}}^{(K)}
\bigr\}
$
is a collection of enhanced feature embeddings where each embedding has been refined by selectively 
integrating context from all other features within the collection. This selective integration captures how certain attributes from one feature can inform 
or modify the representation of another, resulting in a more comprehensive and powerful fused representation of the command features. Section \ref{Interpretability} provides visualizations of learned attention weights. 

We then employ an attentive pooling inspired by \cite{wu-etal-2020-attentive} to summarize the collection of fused features $\widehat{\mathbf{X}}_{c_t^{(s)}} \in \mathbb{R}^{K \times D}$. Let \(\mathbf{q} \in \mathbb{R}^{1 \times D}\) be a trainable query vector. We compute an attention score for each fused feature \(\widehat{\mathbf{x}}_{c_{t}^{(s)}}^{(k)} \in \widehat{\mathbf{X}}_{c_t^{(s)}}\)  by projecting it onto \(\mathbf{q}\), then normalize these scores with softmax:
\begin{equation}
\label{eq:att_score}
\alpha_{t,k} 
\;=\;
\frac{\exp\!\bigl(\widehat{\mathbf{x}}_{c_{t}^{(s)}}^{(k)} \cdot \mathbf{q}\bigr)}{%
\sum_{j=1}^{K}\exp\!\bigl(\widehat{\mathbf{x}}_{c_{t}^{(s)}}^{(j)} \cdot \mathbf{q}\bigr)}
\end{equation}
where \(\alpha_{t,k}\) is the attention weight for each feature type \(k\in \widehat{\mathbf{X}}_{c_t^{(s)}}\). A final weighted sum produces a fused representation \(\mathbf{z}_t \in \mathbb{R}^{1 \times D} \) that integrates and summarizes all available meta-information for command \(c_{t}^{(s)}\). 
\begin{equation}
\label{eq:weight_sum}
\mathbf{z}_t
\;=\;
\sum_{k=1}^{K} 
\alpha_{t,k}\,\widehat{\mathbf{x}}_{c_{t}^{(s)}}^{(k)}
\end{equation}
This attention-based fusion enriches the context at each time step by learning how different feature types contribute to the overall command representation.

We then collect these fused command embeddings \(\mathbf{z}_1, \mathbf{z}_2, \ldots, \mathbf{z}_n\) along the session to form a sequence-level encoding of the input command sequence. This sequence will be fed into the subsequent Transformer backbones, which learn the temporal and contextual relationships among commands within the session.
By separately handling feature fusion within each command (intra-command), and sequence modeling across commands (inter-command) through Transformers, our model not only captures the nuanced details of individual commands but also tracks how the session evolves over time, thereby improving the accuracy of next-command predictions.

\subsubsection{Masking strategies}
\label{sec:mask}

Before feeding the command embedding sequence into the Transformer backbone, the embedding corresponding to the command to be predicted in each sequence is masked, while the associated command information (e.g., ID) is stored as ground truth labels to enable self-supervised training. In this work, we adapt two primary masking strategies from NLP: Causal Language Modeling (CLM) \cite{radford2018improving} and Masked Language Modeling (MLM) \cite{devlin-etal-2019-bert}. Unlike in NLP, which often represents masks with a special token, we use a learnable mask embedding to better align with our use case.

CLM masks the right side of the sequence, predicting the next item based on its preceding context. This setup prevents the model from accessing future items, making it well suited for decoder-only Transformer architectures such as GPT \cite{gpt-2}, as it aligns with their commonly employed causal attention mechanism.

On the other hand, MLM randomly masks approximately 15\% of the items throughout the sequence and allows the model to leverage the bidirectional context when predicting these masked items. This strategy is employed in encoder-only Transformer architectures such as BERT \cite{devlin-etal-2019-bert} and encoder-decoder architectures such as T5 \cite{T5}, as the self-attention mechanisms in the Transformer encoder can access bidirectional context.

Although MLM enables the model to incorporate future context during training, exposing future information during inference is incompatible with the objective of next-command recommendation as defined in Section \ref{problem_statement}. Consequently, regardless of the masking strategy, we evaluate model performance during inference by masking only the last command in the sequence.

\subsubsection{Transformer backbones}
\label{transformer_backbone}

In this section, we introduce different Transformer backbones utilized in the proposed model. To leverage the strengths of large language models (LLMs) in sequence representation learning and to explore their transferability to the command recommendation task, this backbone stacks $N$ Transformer blocks derived from the latest LLM architectures. Specifically, we systematically compare four common structures: Encoder-only, Decoder-only, Decoder-only with MoE, and Encoder-Decoder. The primary function of the Transformer backbone is to represent the input sequence as context-aware embeddings and hidden states, thereby capturing the global dependencies across different positions within the input sequence.

\begin{figure}[ht!]
    \begin{subfigure}[b]{0.5\textwidth}
        \centering
        \includegraphics[width=0.7\textwidth]{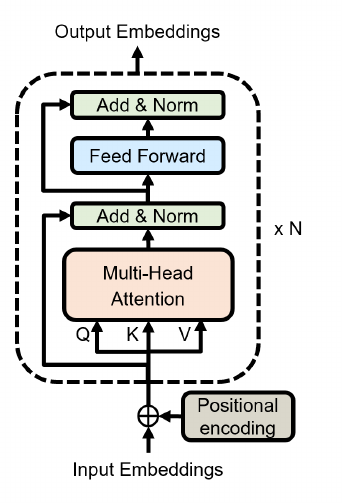}
        \caption{Encoder-only (BERT)}
        \label{fig:sub1}
    \end{subfigure}
    \begin{subfigure}[b]{0.5\textwidth}
        \centering
        \includegraphics[width=0.8\textwidth]{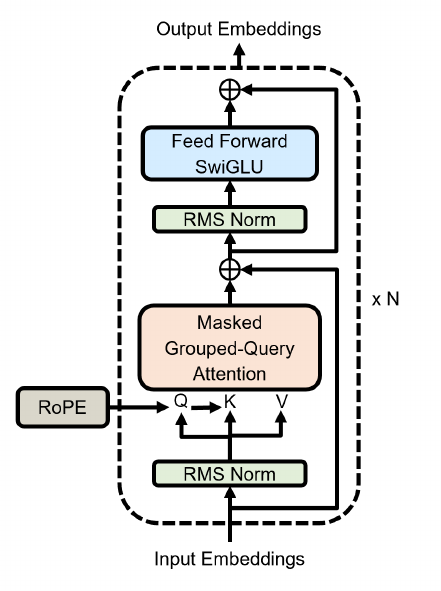}
        \caption{Decoder-only (Llama3.2)}
        \label{fig:sub2}
    \end{subfigure}
    \begin{subfigure}[b]{0.5\textwidth}
        \centering
        \includegraphics[width=\textwidth]{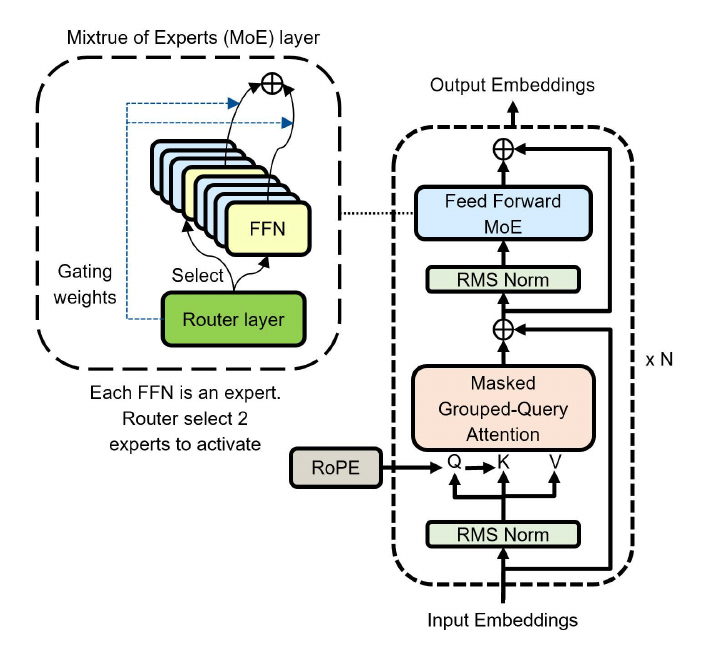}
        \caption{Decoder-only with MoE (Mixtral)}
        \label{fig:sub3}
    \end{subfigure}
    \begin{subfigure}[b]{0.5\textwidth}
        \centering
        \includegraphics[width=\textwidth]{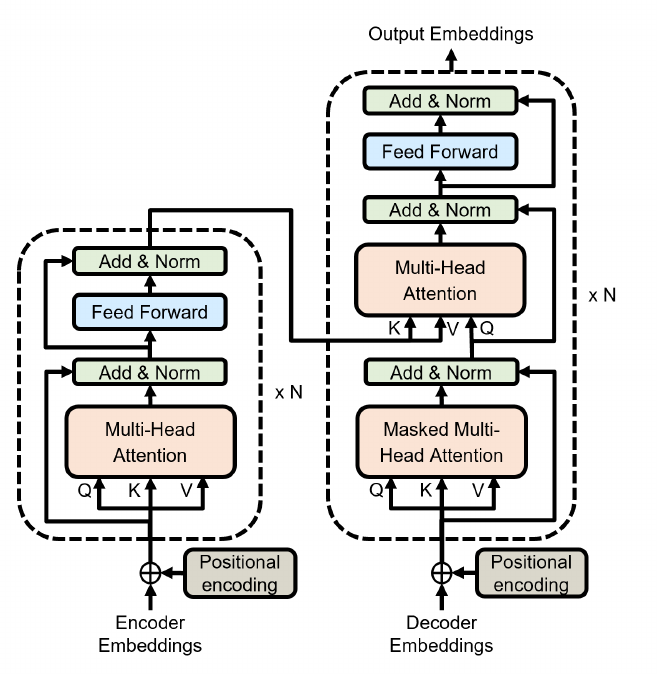}
        \caption{Encoder-Decoder (T5)}
        \label{fig:sub4}
    \end{subfigure}
    
    \caption{Different Transformer backbones used in this study}
    \label{fig:main}
\end{figure}

\textbf{Encoder-Only.} We adopt the Transformer blocks from BERT \cite{devlin-etal-2019-bert} because of its representative encoder-only architecture. As shown in Figure~\ref{fig:sub1}, each block (or layer) primarily consists of a Multi-Head Self-Attention and a Feed-Forward Network (FFN). Given an input embedding sequence \( S \in \mathbb{R}^{n \times d} \), where \( n \) represents the sequence length and \( d \) is the feature dimension, the input embeddings are first combined with positional encodings. These embeddings are then linearly projected to generate three tensors: \( Q \) (Query), \( K \) (Key), and \( V \) (Value). The attention scores are computed using the scaled dot-product attention formula \cite{transformer}, which can be expressed as:

\begin{equation}
\label{eq:att_form}
\text{Attention}(Q, K, V) = \text{softmax}\left(\frac{QK^\top}{\sqrt{d_k}}\right)V
\end{equation}

Here, \( d_k \) is the dimension of the query and key in a single attention head. The multi-head attention extends this by performing multiple parallel attention computations, enabling the model to capture richer feature representations. The outputs of all attention heads are concatenated along the feature dimension and then linearly transformed back to the original size. The output of the multi-head attention module is added to the input embeddings via a residual connection, followed by layer normalization \cite{DBLP:journals/corr/BaKH16}.
The normalized result is then passed through the FFN, which consists of two fully connected layers with a GELU activation function \cite{hendrycks2023gaussianerrorlinearunits} in between. The output of the FFN is again combined with the residual connection from its input and normalized, producing an output \( S_{\text{out}} \in \mathbb{R}^{n \times d} \), which matches the shape of the input embeddings.
These BERT-style Transformer encoder blocks are stacked \( N \) times to form the backbone network. A notable characteristic of the encoder is that it does not apply masking when computing \( Q, K, V \), thereby enabling the model to consider bidirectional context over the entire input sequence.

\textbf{Decoder-Only.} The architecture adopted by most mainstream LLMs, such as GPT \cite{gpt-2}, Llama \cite{grattafiori2024llama3herdmodels}, and Mistral \cite{jiang2023mistral7b}, focuses on unidirectional context modeling, meaning that only items before the current item are taken into account for the prediction. This is required for tasks like text generation as the preceding text is not known at deployment time. Figure~\ref{fig:sub2} illustrates the Transformer blocks of the representative Llama3.2 model. At a high level, its processing of input sequence embeddings is similar to that of an encoder-only approach. However, the self-attention as illustrated in Eq.~\ref{eq:att_form} is masked by causal masking (causal self-attention mechanism), meaning that at the current item position, the query only attends to the hidden states of current and previous items.
As the state-of-the-art LLM, Llama3.2 \cite{grattafiori2024llama3herdmodels} introduces several improvements to the native Transformer decoder architecture. These include replacing absolute positional embeddings with rotary positional embeddings (RoPE) \cite{su2023roformerenhancedtransformerrotary}, adopting root mean square normalization (RMSNorm) \cite{rmsnorm} instead of layer normalization \cite{DBLP:journals/corr/BaKH16}, employing the SwiGLU activation function \cite{shazeer2020gluvariantsimprovetransformer}, and leveraging grouped-query attention (GQA) to reduce the memory footprint of attention computation and improve inference speed \cite{grattafiori2024llama3herdmodels}.

\textbf{Decoder-only with MoE.} The concept of Mixture of Experts (MoE) is designed to create a regulatory framework for systems composed of multiple individual networks. In such systems, each network, or "expert", is responsible for processing a specific subset of inputs, concentrating on a particular region of the input space. A Router network determines which experts handle a given input by assigning weights to each expert and combining their outputs accordingly \cite{sanseviero2023mixture}, as illustrated in Figure~\ref{fig:sub3}. In our study, we utilized Transformer blocks derived from Mixtral \cite{jiang2024mixtralexperts}, which share a similar structure to Llama but replace the dense Feed Forward Network (FFN) layers with sparse MoE layers. Each MoE layer consists of 8 experts, where each expert is a small FFN. During inference, the router layer activates two of these experts based on learned parameters to process the input. Since only a subset of model parameters is activated during inference, MoE models require less computation compared to dense models with the same parameter amount, enabling faster inference.

\textbf{Encoder-Decoder.} This structure integrates a bidirectional encoder and a causally masked decoder. During decoding, cross-attention \cite{transformer} is added to incorporate the encoder's outputs by treating them as key and value. In our model, we stack Transformer blocks derived from the representative T5 model \cite{T5}, as illustrated in Figure~\ref{fig:sub4}, whose structure is close to the naive Transformer. We employ the MLM method mentioned in Section~\ref{sec:mask}, randomly masking 15\% of the items in the encoder input sequence, allowing the model to learn to predict and complete them. To construct the target sequence, we retrieve the corresponding embeddings of the labels to be predicted from the embedding table and shift them rightward by one position to form the decoder's input embeddings. This enables training of such seq2seq models using the teacher forcing approach \cite{10.5555/3157382.3157612}.

\subsubsection{Multi-task learning and loss function}

Multi-task learning has been widely applied across various domains due to its ability to reduce overfitting and improve model generalization by leveraging complementary information from multiple objectives \cite{wang2023multitaskdeeprecommendersystems}. Specifically, multi-task learning encourages the model to learn shared representations that are meaningful across all tasks, enhancing the overall learning process through inter-task reinforcement. It often proves beneficial when a primary task can be supported by auxiliary objectives that convey relevant domain knowledge \cite{zhang2023advanceschallengesmultitasklearning}. Inspired by this, we enhance the primary task of the next-command ID recommendation by jointly learning two auxiliary classification tasks (command type and target prediction). 

Specifically, the output of the Transformer backbone is fed into three separate fully connected (FC) branches (prediction heads), each projecting into a distinct output space corresponding to three different tasks. The outputs from these branches are processed using softmax to generate probability distributions over their respective label spaces. Considering the imbalance and long-tail distribution commonly observed in the command patterns within BIM logs, we adopt the focal loss \cite{focal_loss} for each task. Focal loss modifies the standard cross-entropy loss by introducing a modulation factor that down-weights the contribution of well-classified examples. This shifts the focus more toward difficult or easily misclassified examples, helping to address class imbalance issues. The focal loss for a single task can be expressed as:

\begin{equation}
\label{eq:focal_loss}
\text{FL}(p_t) = 
- \alpha_t (1 - p_t)^{\gamma} \log (p_t),
\end{equation}
where $p_t$ is the model's estimated probability for the ground-truth class, $\alpha_t$ is a weighting factor to address class imbalance, and $\gamma$ is a focusing parameter that smoothly adjusts the rate at which easy examples are down-weighted. Let the losses for the three tasks be $\ell_{\text{cmd}}, \ell_{\text{typ}}, \ell_{\text{tgt}}$, the overall training objective $\ell_{\text{total}}$ is then a weighted sum of these losses:
\begin{equation}
\label{eq:total_loss}
\ell_{\text{total}} 
= \ell_{\text{cmd}} + \gamma_{1} \, \ell_{\text{typ}} + \gamma_{2} \, \ell_{\text{tgt}},
\end{equation}
where $\gamma_{1}$ and $\gamma_{2}$ are hyperparameters controlling the relative importance of the auxiliary tasks. 

During backpropagation, gradients from all three tasks flow through the shared Transformer backbone, allowing the model to learn richer and more generalizable representations. By incorporating auxiliary objectives and focal loss, our model more effectively handles imbalanced data and captures diverse signals that are relevant for robust sequential recommendation. 
As stated in Section \ref{problem_statement}, the model ultimately outputs a probability distribution over all possible next-step commands, from which we select the top K commands as the recommendation list.

\newpage
\section{Case study}
\label{case study}

\subsection{Dataset}

\begin{figure}[ht]
    \centering
    \includegraphics[width=\linewidth, clip, trim=5mm 60mm 5mm 3mm]{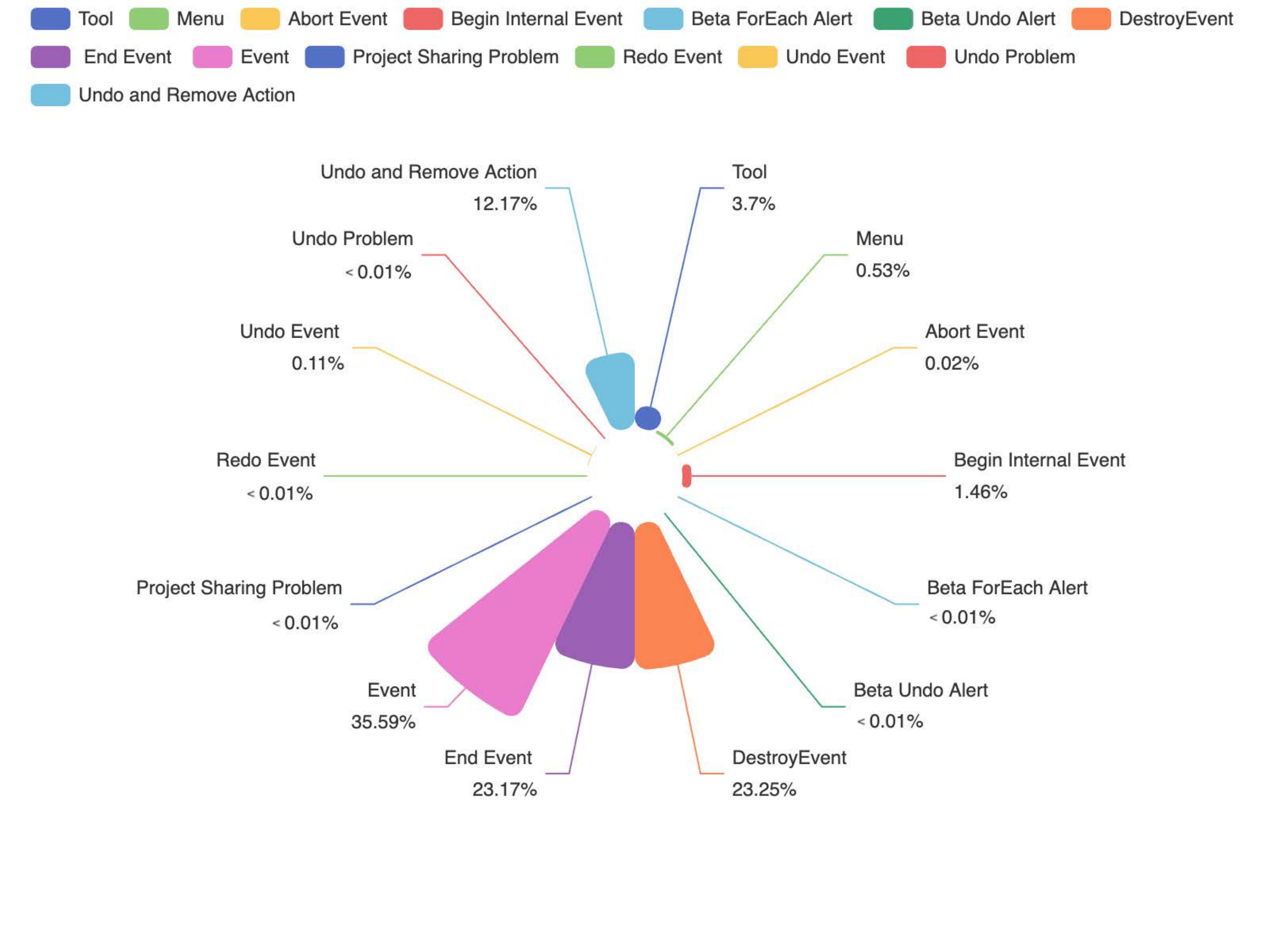}
    \caption{Percentage distribution of commands by prefix in the raw dataset.}
    \label{Percentage}
\end{figure}

In the presented case study, we utilized real-world, large-scale BIM log data from Vectorworks, which comprises a total of 630 parquet files compressed into 21 zipped files, approximately 90GB in size. The log dataset represents six months of anonymous user activity collected across the world, totaling 32,374,984,579 log entries, spanning 7 languages, 1,658,525 sessions, and 97,452 unique command instances. Figure~\ref{fig:native_logs} shows an example log snippet. 
The commands are pre-categorized as UNDO, Tool, and Menu, which are further classified into 14 distinct prefixes defined in the message column. Figure \ref{Percentage} demonstrates the command distribution based on the prefix. Detailed explanations and numerical breakdowns of each category/prefix are provided in Table \ref{table_categories}. 

Among various prefixes, \textit{Tool} and \textit{Menu} typically record high-level commands accessible through the UI (such as button names). In contrast, frequently occurring commands with prefixes like \textit{Event} and \textit{End Event} generally represent the smallest operational units recorded in the Vectorworks log, as shown in Figure \ref{fig:native_logs}. Further analysis reveals that these "event-related" commands dominate for two main reasons. First, during modeling, cursor actions (such as dragging and zooming) and keyboard shortcuts frequently trigger these commands directly. Second, high-level UI-accessible commands like \textit{Tool} and \textit{Menu}, trigger multiple low-level commands (events) to log their initiation, interruption, and completion. This behavior results in multiple command entries capturing the same user action, leading to redundancy in log data. However, due to the lack of explicit rules and documentation to systematically define these triggering relationships, effectively filtering out true user actions remains a challenge.

Each command in a BIM log is usually associated with a unique ID \cite{JANG2023102079}. However, analysis of the log dataset reveals that, in some cases, command IDs are not strictly correlated with command names. For instance, as shown in the first column of Table \ref{tab:language_dic}, different commands may share the same ID. This inconsistency, combined with the multilingual nature, significantly increases the complexity of the dataset, necessitating an effective data processing approach to standardize command representations.

\subsection{Data processing}
\label{data_processing}

The raw log dataset is processed following the proposed multi-stage data filtering and enhancement method as illustrated in Section \ref{datafilteringworkflow}. 

\textbf{First}, to accurately track the actual command flow, we adhere to the principles outlined in Section \ref{actualmodelingflowtracking} to systematically filter out irrelevant commands, including internal software events/alerts, low-significance actions, and aborted or unfinished commands. Subsequently, Algorithm \ref{algorithm_undo_redo} is applied to each session to remove undone commands while restoring those that have been redone.


\textbf{Second}, to address the challenge of handling multiple languages, Algorithm \ref{algorithm_language} as outlined in Section \ref{multilanguagealignment} is applied. Although various translation services are available, we chose the widely recognized and commonly used Google Translate API \cite{GoogleTranslate} to translate the distinct commands into English. To further standardize and align these translations, we employed the state-of-the-art text embedding model text-embedding-3-large \cite{openai2024embedding} to generate semantic embeddings. An empirical threshold of 0.82 is set to evaluate the semantic similarity of the command names within each ID group. For groups smaller than this threshold, DBSCAN \cite{DBSCAN1} is applied for further clustering. Bayesian optimization \cite{10.5555/2999325.2999464} is used to automatically determine the optimal value of the $\varepsilon$-neighborhood radius in DBSCAN, with the objective of maximizing the Silhouette Score \cite{ROUSSEEUW198753}, an unsupervised metric for assessing clustering quality.
Table \ref{tab:language_dic} illustrates examples comparing raw commands, translated commands, and the final aligned commands outputted from this module. By combining with embedding-based clustering, the multilingual commands were successfully aligned.

\begin{CJK}{UTF8}{}
\begin{table}[ht!]
\caption{Example outcomes of the multi-language alignment. In the raw log data, "Create Roof (166)" and "Create Line (166)" share the same command ID (166) while "Create Object (92)" is consistently associated with ID 92 across all instances. The column "Translated commands" highlights biases introduced by the translation API, including inconsistencies in letter case and synonymous terms. After applying the alignment process, these commands are standardized with a unified representation in English, which replaces the unreliable ID to become a unique representation of each command.}
\centering
{\footnotesize
\begin{tabular}{lcc}
\toprule
\textbf{Multilingual commands}      & \textbf{Translated commands} & \textbf{Aligned commands} \\
\midrule
Create Object (92) & Create Object (92) & Create Object \\
Objekt anlegen (92) &Create object (92) & Create Object \\
Crea Oggetto (92)& Create Object (92)& Create Object \\
Crear objeto (92) & Create object (92) & Create Object\\
Criar Objeto (92) & Create Object (92) & Create Object \\
\zh{创建对象} (92) & Create Object (92) & Create Object \\
\jp{図形の生成} (92) & Shape creation (92) & Create Object \\

\midrule
Create Roof (166)               & Create Roof (166)        & Create Roof       \\
Dach anlegen (166)               & Create roof (166)	       & Create Roof        \\
Creëer dak (166)               & Create roof (166)	        & Create Roof       \\
Utwórz dach (166)              & Create Roof (166)	        & Create Roof        \\
Crea Tetto (166)            & Roof creation (166)	        & Create Roof \\
Gerar Telhado (166)           & Generate Roof (166)        & Create Roof \\
\jp{屋根作成} (166)            & Roof creation (166)	        & Create Roof \\

\midrule
Create Line (166)           & Create Line (166)	        & Create Line \\
Linie anlegen (166)          & Create Line (166)	        & Create Line \\
Creëer lijn (166)          & Create Line (166)	        & Create Line \\
Créer une ligne (166)          & create a line (166)	        & Create Line \\
Crear línea (166)           & Create line (166)	        & Create Line \\
\zh{创建线条} (166)           & Create Line(166)	        & Create Line \\
\jp{線分の生成}(166)          & Line creation (166)	        & Create Line\\

\bottomrule
\end{tabular}
}
\label{tab:language_dic}
\end{table}
\end{CJK}

\begin{figure}[ht!]
    \centering
    \includegraphics[width=1\linewidth, clip, trim=8mm 214mm 8mm 9.5mm]{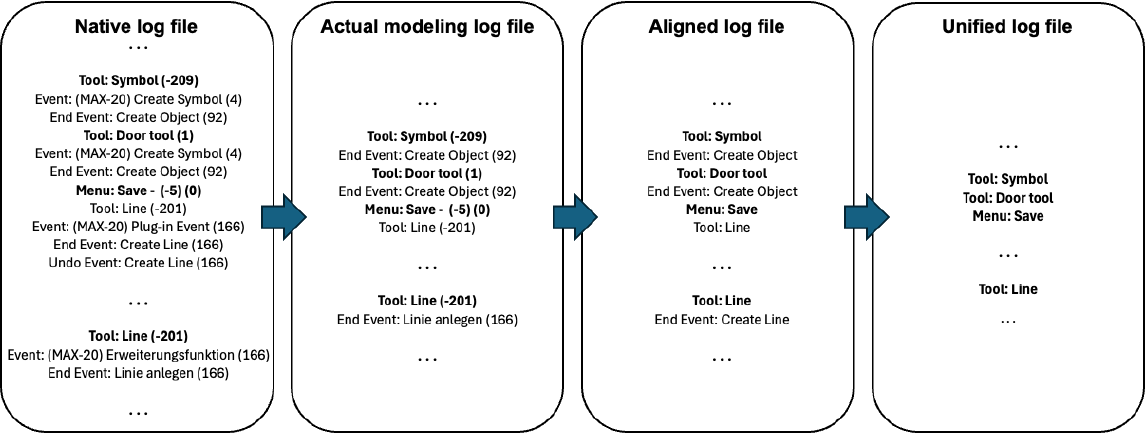}
    \caption{Visualization of the log data filtering from the native log file to the unified log file, demonstrating example outcomes from each module illustrated in Section \ref{datafilteringworkflow}. The commands in bold belong to the users' true action flows.}
    \label{log_compare}
\end{figure}

\textbf{Third}, to identify and remove redundant commands in log data, as outlined in Section \ref{redundantcommand}, the primary objective is to remove low-level commands triggered by high-level commands by mining the mapping between them. A confidence threshold of 0.4 is used in Association Rule Mining (ARM) \cite{arm}, below which no significant relationship between commands is assumed. For the commands exceeding the threshold, the top 10 most likely low-level commands triggered by the high-level command are identified. These triggering behaviors are further manually validated in Vectorworks. Based on the mined command mappings, only the successfully completed high-level commands (i.e. those that successfully triggered and completed the corresponding low-level commands, if any) are retained to represent each action. Figure \ref{log_compare} visualizes the process of log data filtering from the native log file (Figure \ref{fig:native_logs}) to the unified log file. The native log initially contains 16 entries representing four essential user actions: creating a symbol, using the door tool, saving, and creating a line in the German version. These actions are performed across two languages and two sessions. Through processing, redundant low-level commands are removed, reducing the log to four concise entries that accurately capture user actions.

\begin{figure}[ht]
    \centering
     \begin{subfigure}[c]{0.49\textwidth}
        \centering
        \includegraphics[width=1\linewidth]{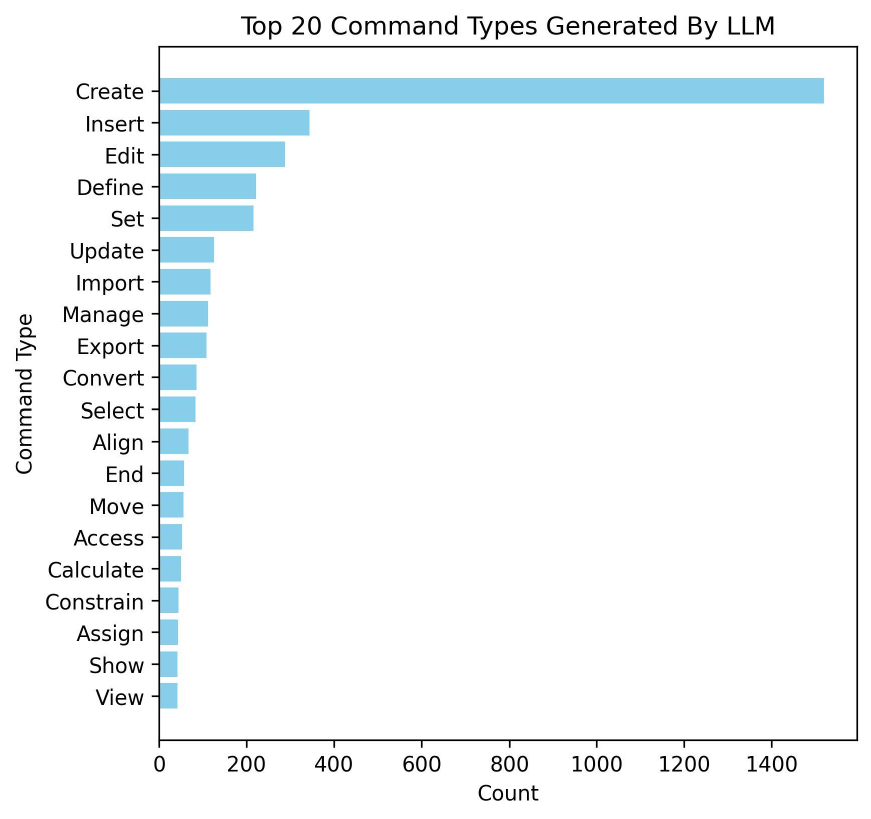}
        \caption{Top 20 command types generated by LLM based on the software documentation}
        \label{top20journal}
    \end{subfigure}
    \begin{subfigure}[c]{0.49\textwidth}
        \centering
        \includegraphics[width=1\linewidth]{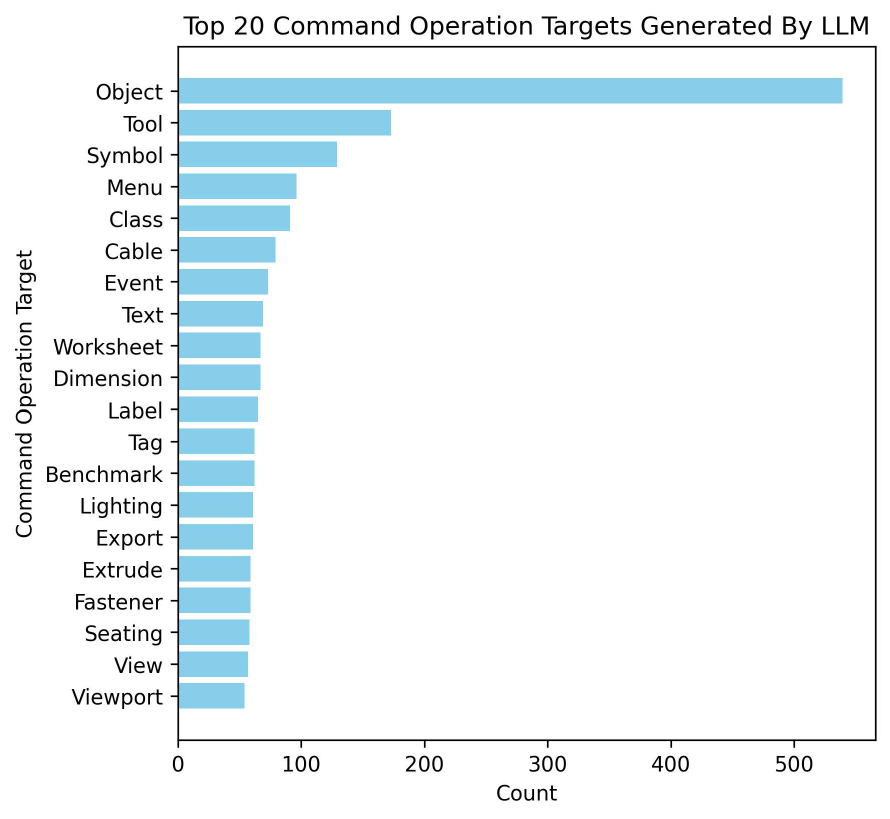}
        \caption{Top 20 command operation targets generated by LLM based on the software documentation}
        \label{top20targets}
    \end{subfigure}
    \caption{Command types and targets generated by the augmentation method}
\end{figure}

\textbf{Fourth}, in the process of information augmentation and workflow generation, as outlined in Section \ref{augmentationandpostprocessing} and Figure \ref{fig:information_augmentation}, we cleaned and extracted 1,911 HTML files from Vectorworks' online documentation into Markdown format. This documentation was processed using the text-embedding-3-large \cite{openai2024embedding} to generate semantic embeddings for a vector database. Following the custom RAG pipeline developed by \cite{Du_Copilot_24}, the two most semantically relevant contents were retrieved from the vector database for each command name. The extracted domain knowledge was summarized by the GPT-4o-mini \cite{GPT4}, generating descriptions for each command. 
We chose GPT-4o-mini because it excels in text summarization while also offering high speed and low cost, making it particularly suitable for processing our large-scale data.
Based on the description information, GPT-4o-mini further categorized the commands into 174 types and inferred 363 possible command operation targets to provide rich meta-information, as illustrated in Figure \ref{top20journal} and Figure \ref{top20targets}.

As detailed in Section \ref{augmentationandpostprocessing}, a modified BPE algorithm is applied to generate 10 common workflows, expanding the command vocabulary to 4939. GPT-4o-mini is further utilized to generate workflow descriptions, types, and operation targets based on the metadata of the commands composing each workflow. Table \ref{merged_table} highlights the example commands and workflows along with their associated additional information generated by LLM. 

In addition to enhancing command metadata using LLMs, we also leverage information inherently present in logs to compute statistical features of commands within each session. These features include the execution time (time interval) of commands and the frequency of consecutive repetitions within the same session. Specifically, we merge temporally consecutive duplicate command entries into a single entry and record the number of consecutive occurrences as a feature.

Given the significant variation in session lengths within the log data (ranging from a few entries to tens of thousands), we exclude sessions with fewer than five interactions. Infrequent commands that appeared fewer than 10 times in the dataset are also removed.
Considering the context window limitations of Transformer models, we set 110 for the maximum sequence length, balancing contextual richness with computational efficiency. Additionally, to simulate the dynamic growth of session lengths in production environments, we sample overlong sessions into randomly sized subsequences ranging from 10 to 110 based on temporal order. This approach significantly increases the number of session (i.e., sequence) samples in the dataset. Table \ref{original_cleaned_dataset} summarizes the differences between the initial raw data and the final dataset after augmentation and processing. 

\begin{table}[ht]
\caption{Statics of the original and final datasets}
\centering
\begin{tabular}{lcc}
\toprule
\textbf{Metric}      & \textbf{Original dataset} & \textbf{Final dataset} \\
\midrule
Entries                   & 32,374,984,579           & 375,304,719              \\
Sessions                  & 1,658,525                & 6,856,392                \\
Average session length                  & 19,520               & 55               \\
Unique commands           & 97,452                   & 4,939 (incl. workflows)                    \\
Types                & None            & 174                      \\
Targets                   & None           & 363                      \\
Descriptions                   & None           & 4,939                      \\
\bottomrule
\end{tabular}

\label{original_cleaned_dataset}
\end{table}

\begin{figure}
    \centering
    \includegraphics[width=1\linewidth]{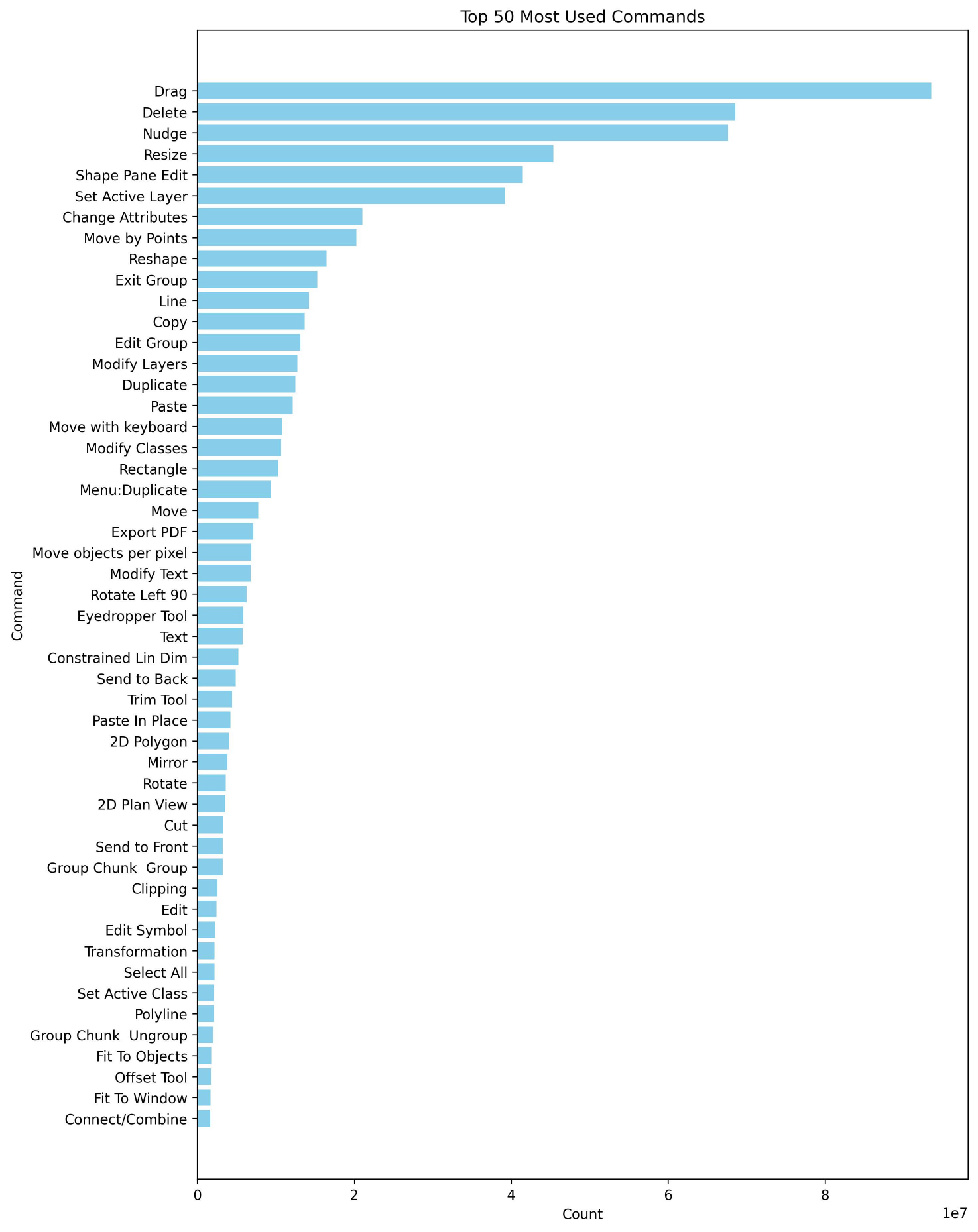}
    \caption{Top 50 most recorded commands in Vectorworks in the final dataset}
    \label{top50}
\end{figure}

Figure \ref{top50} shows the 50 most frequently logged commands in the final dataset.
The dataset demonstrates an imbalanced long-tailed behavior, where a subset of commands occurs with very high frequencies. These commands often represent shortcut/cursor actions, such as \textit{Drag}, \textit{Resize}, and \textit{Delete}, which are frequently used by users for small, rapid modifications to the BIM model.



\newpage
\section{Experiments and results}

All data processing and experiments were conducted on a workstation with a 48GB Quadro RTX 8000 GPU and 250GB RAM. We randomly sample 85\% of the sessions from the final dataset as the training set and 15\% as the validation set (resulting in 1,031,839 test samples), while ensuring that both sets contain all unique commands. The session (sequence) lengths in both sets vary dynamically from 5 to 110 to best simulate real-world production environments. The validation follows the standard approach in sequential recommendation: only the last command in the input sequence is masked, and the model's performance is evaluated based on its ability to predict the masked item given the preceding context.

Given the large dataset, all models were trained for 10 epochs with a batch size of 128, averaging 42 hours per training run. We used the Adam optimizer with an initial learning rate of 3e-5, which linearly decayed with training steps. Due to GPU hardware limitations, we preserved the original configurations of different Transformer blocks (e.g., number of attention heads, hidden layer dimensions, etc.) as much as possible. To ensure the model size remained compatible with our hardware, we only stacked two Transformer blocks in our model.

\subsection{Evaluation metrics}

In this study, we employ the Recall@K and NDCG@K metrics, which are widely used in sequential recommendation \cite{BOKA2024102427}, to assess the recommendation quality of the proposed model. The necessary presumptions for these two evaluation metrics are discussed in Section \ref{evaluation_strategy}.

\textbf{Recall@K} measures the proportion of test instances where the ground-truth next item appears among the top-K recommended items. Formally, if \( R \) represents the top-K recommendations for a given test instance and \( y \) is the ground-truth item, then Recall@K is defined as:

\begin{equation}
\text{Recall@K} = \frac{1}{N} \sum_{i=1}^{N} I(y_i \in R_i)
\end{equation}

where \( N \) is the total number of test instances, and \( I(\cdot) \) is an indicator function that returns 1 if the ground-truth item \( y_i \) is present in the top-K recommendations \( R_i \), and 0 otherwise. A higher Recall@K indicates that the model is more effective in retrieving the correct item within the top-K recommendation list.

\textbf{NDCG@K} (Normalized Discounted Cumulative Gain) \cite{10.1145/582415.582418} considers the ranking positions of relevant items in the top-k recommendation list. It is a normalized version of Discounted Cumulative Gain (DCG), which applies a logarithmic discount factor to reduce the contribution of relevant items that appear lower in the ranking. For a single relevant item at position \( p \) in the top-K list, DCG is computed as:

\begin{equation}
DCG = \frac{1}{\log_2 (p+1)}
\end{equation}

NDCG normalizes this score by the maximum possible DCG, i.e., the ideal DCG (\( IDCG \)) when the relevant item appears in the first position. Thus, NDCG@K is defined as:

\begin{equation}
NDCG@K = \frac{1}{N} \sum_{i=1}^{N} \frac{DCG_i}{IDCG}
\end{equation}

A higher NDCG@K indicates that relevant items tend to appear closer to the top of the ranked list, reflecting better ranking quality.

Since the goal in a production environment is typically to provide users with the top recommendations rather than the full probability distribution of all commands, we consider setting K to 3, 5, or 10 to be reasonable.

\subsection{Comparison of Transformer architectures}




Table \ref{compare_transformer} compares the performance of our model when using different Transformer backbones. For each type of backbone network, we removed the proposed feature fusion, multi-task learning, and focal loss modules to establish a baseline for comparison. The resulting baseline models are similar to the settings used in the previous study \cite{GAO2022104026}.

Overall, our proposed model architecture outperforms the baselines across all metrics, regardless of the Transformer backbone used. 
We further employed the paired bootstrap method to compare the performance of our proposed model with the baseline. The calculated 95\% confidence intervals and p-values demonstrate that the improvement is statistically significant. Detailed results can be found in \ref{app:statis}.
This also demonstrates that the augmented features introduced in Section \ref{augmentationandpostprocessing} have a generalizable positive effect on improving command recommendation. 
Comparing different backbone networks, decoder-only architectures such as Mixtral-MoE and Llama3.2 outperform BERT and T5 due to their more recent designs and advanced techniques, despite the latter two having the ability to learn bidirectional contextual information. Among them, Mixtral-MoE achieves the best performance, with the correct next-step command appearing in the top-5 and top-10 recommendations with approximately 71.7\% and 83.3\% probability, respectively.

\begin{table}[ht]
\centering
\footnotesize
\caption{Comparison of Transformer architectures used in the model}
\renewcommand{\arraystretch}{1.2} 
\setlength{\tabcolsep}{2pt}      
\label{compare_transformer}
\begin{tabular}{%
|>{\centering\arraybackslash}p{2.7cm}%
|>{\centering\arraybackslash}p{2.2cm}%
|>{\centering\arraybackslash}p{1.8cm}%
|>{\centering\arraybackslash}p{1.2cm}%
|>{\centering\arraybackslash}p{1.2cm}%
|>{\centering\arraybackslash}p{1.2cm}%
|>{\centering\arraybackslash}p{1.2cm}%
|>{\centering\arraybackslash}p{1.2cm}%
|>{\centering\arraybackslash}p{1.2cm}|}

\hline
\textbf{Transformer type} & \textbf{Transformer backbone\textsuperscript{*}} & \textbf{Method\textsuperscript{**}} & \textbf{Recall @3} & \textbf{NDCG @3} & \textbf{Recall @5} & \textbf{NDCG @5} & \textbf{Recall @10} & \textbf{NDCG @10} \\ 
\hline
\multirow{2}{*}{Encoder-only} 
 & \multirow{2}{*}{BERT} 
 & Proposed & 60.299 & 50.607 & 71.071 & 55.047 & 83.025 & 58.942 \\ \cline{3-9}
 &  & Baseline & 59.925 & 50.193 & 70.766 & 54.658 & 82.923 & 58.619 \\ \hline\hline
\multirow{4}{*}{Decoder-only} 
 & \multirow{2}{*}{Mixtral-MoE} 
 & Proposed & \textbf{61.252}  & \textbf{51.620}  & \textbf{71.709}  & \textbf{55.930} & \textbf{83.307} & \textbf{59.710} \\ \cline{3-9}
 &  & Baseline & 60.021 & 50.354 & 70.739 & 54.77 & 82.726 & 58.678 \\ 
 \hhline{|~*{8}{=}|}
 & \multirow{2}{*}{Llama3.2} 
 & Proposed & 61.112 & 51.480 & 71.612 & 55.807 & 83.229 & 59.594 \\ \cline{3-9}
 &  & Baseline & 59.945 & 50.337 & 70.782 & 54.803 & 82.721 & 58.694 \\ \hline\hline
\multirow{2}{*}{Encoder-Decoder} 
 & \multirow{2}{*}{T5} 
 & Proposed & 60.025 & 50.327 & 70.818 & 54.775 & 82.830 & 58.689 \\ \cline{3-9}
 &  & Baseline & 59.547 & 49.838 & 70.436 & 54.342 & 82.585 & 58.284 \\ \hline
\end{tabular}

\begin{flushleft}
\scriptsize
\textbf{*} For computational efficiency, we stack 2 Transformer blocks as the backbone. \\
\textbf{**} \textit{Proposed} refers to the model architecture introduced in our study, while \textit{Baseline} denotes the setting that excludes modules proposed in this work and uses only the command ID as input to the Transformer (similar to \cite{GAO2022104026}) \\
\end{flushleft}

\end{table}

\subsection{Impact of model size}

\begin{figure}[ht!]
     \begin{subfigure}[b]{0.5\linewidth}
        \centering
         \includegraphics[width=\linewidth]{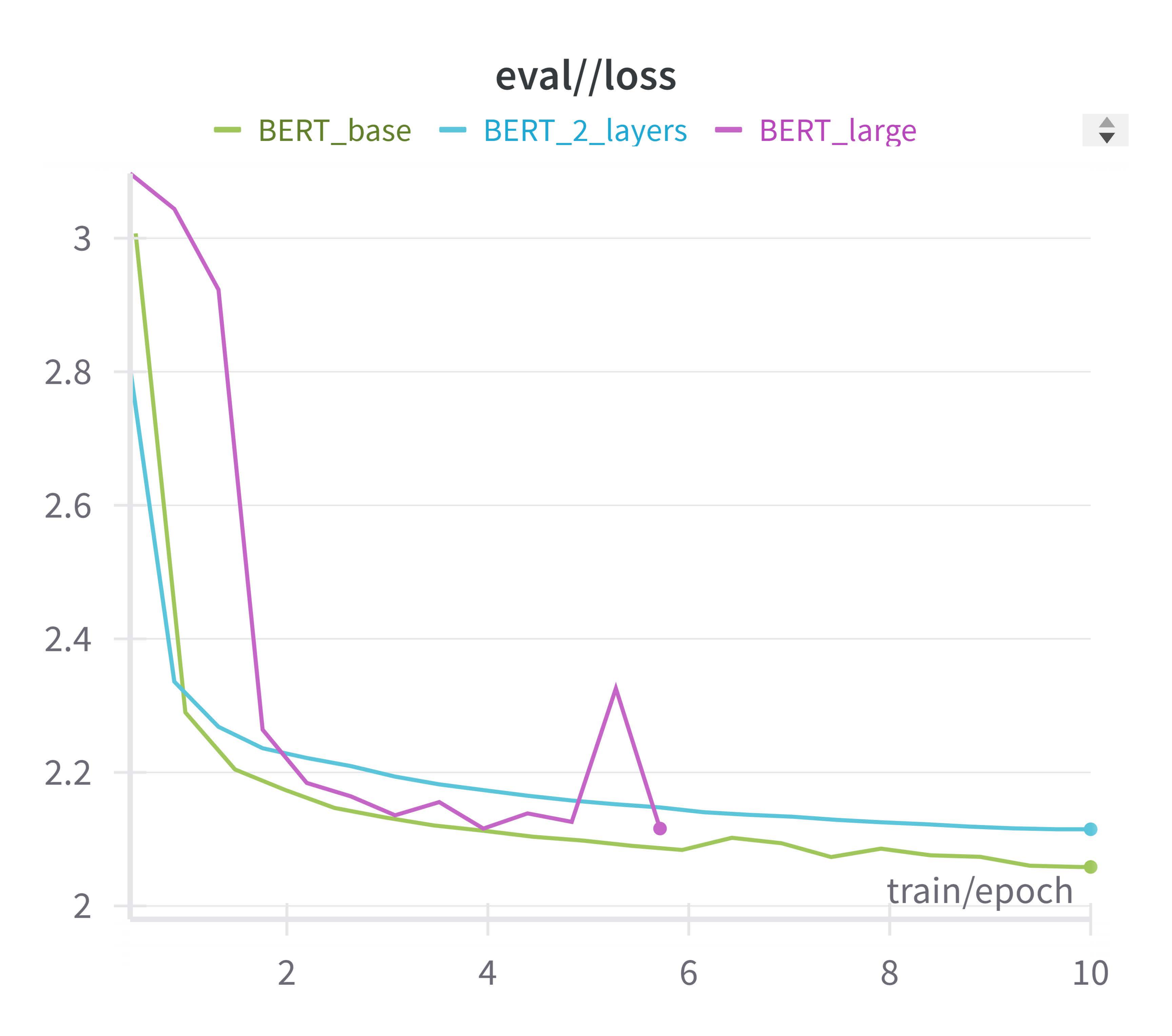}
        \caption{Validation loss}
        \label{fig:bert_loss}
    \end{subfigure}
    \begin{subfigure}[b]{0.5\linewidth}
        \centering
        \includegraphics[width=\linewidth]{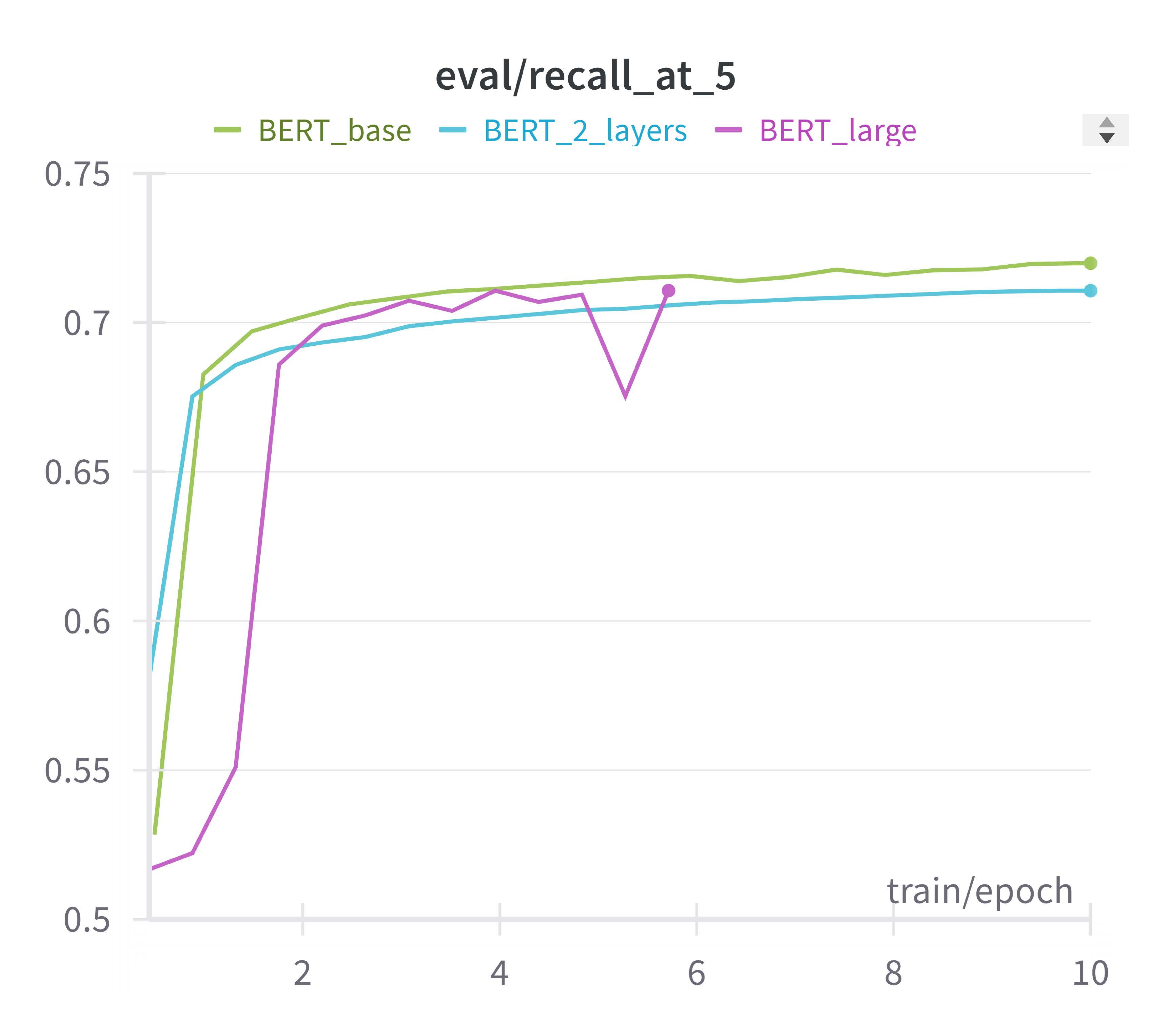}
        \caption{Validation Recall@5}
        \label{fig:bert_recall}
    \end{subfigure}

 \caption{Comparison between models with different sizes of BERT backbone}
\label{model_size}
\end{figure}

In this experiment, we investigate whether increasing the model size improves performance. We chose BERT as the backbone network because, compared to other Transformers, it is relatively smaller and feasible on our hardware. Table \ref{tabel_model_size} presents the model's performance on the validation set when using different sizes of the BERT backbone. Figure \ref{model_size} compares the trends of loss and recall on the validation set as training epochs increase for various model sizes. 
The results show that moderately increasing the model size improves validation metrics; for example, the model using \textbf{BERT-base} achieves a Recall@10 of 83.66\%, surpassing the Mixtral backbone from Table \ref{compare_transformer}. However, excessively large models, such as \textbf{BERT-large}, exhibit unstable training and are prone to overfitting, triggering our early stopping mechanism at epoch 6. This instability may be caused by smaller batches due to hardware limitations or suboptimal learning rates.

\begin{table}[ht!]
\centering
\footnotesize
\caption{Comparing models with different sizes of BERT backbones}
\renewcommand{\arraystretch}{1.2} 
\setlength{\tabcolsep}{2pt}      
\label{tabel_model_size}
\begin{tabular}{%
>{\centering\arraybackslash}p{2.0cm}%
>{\centering\arraybackslash}p{2.8cm}%
>{\centering\arraybackslash}p{1.2cm}%
>{\centering\arraybackslash}p{1.2cm}%
>{\centering\arraybackslash}p{1.2cm}%
>{\centering\arraybackslash}p{1.2cm}%
>{\centering\arraybackslash}p{1.2cm}%
>{\centering\arraybackslash}p{1.2cm}%
>{\centering\arraybackslash}p{1.2cm}}
\hline
\textbf{Transformer backbone*} & \textbf{Setting} & \textbf{Model Size} & \textbf{Recall @3} & \textbf{NDCG @3} & \textbf{Recall @5} & \textbf{NDCG @5} & \textbf{Recall @10} & \textbf{NDCG @10} \\ 
\hline
BERT & 12 heads,2 layers & 53M & 60.299 & 50.607 & 71.071 & 55.047 & 83.025 & 58.942 \\ 
BERT-base & 12 heads,12 layers & 113M & \textbf{61.485} & \textbf{51.819} & \textbf{71.991} & \textbf{56.147} & \textbf{83.663} & \textbf{59.953} \\ 
BERT-large & 16 heads,24 layers & 335M & 60.426 & 50.786 & 71.072 & 55.172 & 82.903 & 59.028 \\ 
\hline
\end{tabular}

\begin{flushleft}
\scriptsize
\textbf{*} We followed the settings of \textbf{bert-base-uncased} and \textbf{bert-large-uncased} from \cite{devlin-etal-2019-bert} and fine-tuned them using their pre-trained weights. \\
\end{flushleft}

\end{table}

\subsection{Ablation study}

To gain a deeper understanding of the contribution of each module in the proposed model, we conducted a detailed ablation study using the Llama3.2 backbone. The results are presented in Table \ref{ablation}. We observed a larger performance drop when all modules were removed, whereas the removal of individual modules led to minor changes. This suggests that while each module's standalone contribution is small, their cumulative effect remains effective. Additionally, considering that the Llama-based baseline model already performs well, even marginal improvements may still be meaningful.

\begin{table}[ht!]
\centering
\footnotesize
\caption{Effectiveness of proposed modules}
\renewcommand{\arraystretch}{1.2} 
\setlength{\tabcolsep}{2pt}      
\label{ablation}
\begin{tabular}{%
>{\centering\arraybackslash}p{5.0cm}%
>{\centering\arraybackslash}p{1.2cm}%
>{\centering\arraybackslash}p{1.2cm}%
>{\centering\arraybackslash}p{1.2cm}%
>{\centering\arraybackslash}p{1.2cm}%
>{\centering\arraybackslash}p{1.2cm}%
>{\centering\arraybackslash}p{1.2cm}}
\hline
\textbf{Methods} & \textbf{Recall @3} & \textbf{NDCG @3} & \textbf{Recall @5} & \textbf{NDCG @5} & \textbf{Recall @10} & \textbf{NDCG @10} \\ 
\hline
Llama3.2 w/ all modules & \textbf{61.112} & \textbf{51.480} & \textbf{71.612} & \textbf{55.807} & \textbf{83.229} & \textbf{59.594} \\ 

w/o att. fusion* & 61.013 & 51.387 & 71.523 & 55.718 & 83.172 & 59.517 \\ 
w/o multi-task & 60.949 & 51.304 & 71.456 & 55.633 & 83.193 & 59.462 \\ 
w/o focal loss & 60.974 & 51.392 & 71.483 & 55.723 & 83.144 & 59.524 \\ 
w/o all modules (baseline) & 59.945 & 50.337 & 70.782 & 54.803 & 82.721 & 58.694 \\ 
\hline
\end{tabular}

\begin{flushleft}
\scriptsize
\textbf{*} Concatenation is used instead of attention-based feature fusion \\
\end{flushleft}

\end{table}

\subsection{Interpretability study}
\label{Interpretability}

\begin{figure}[ht]
    \centering
     \includegraphics[width=1\linewidth]{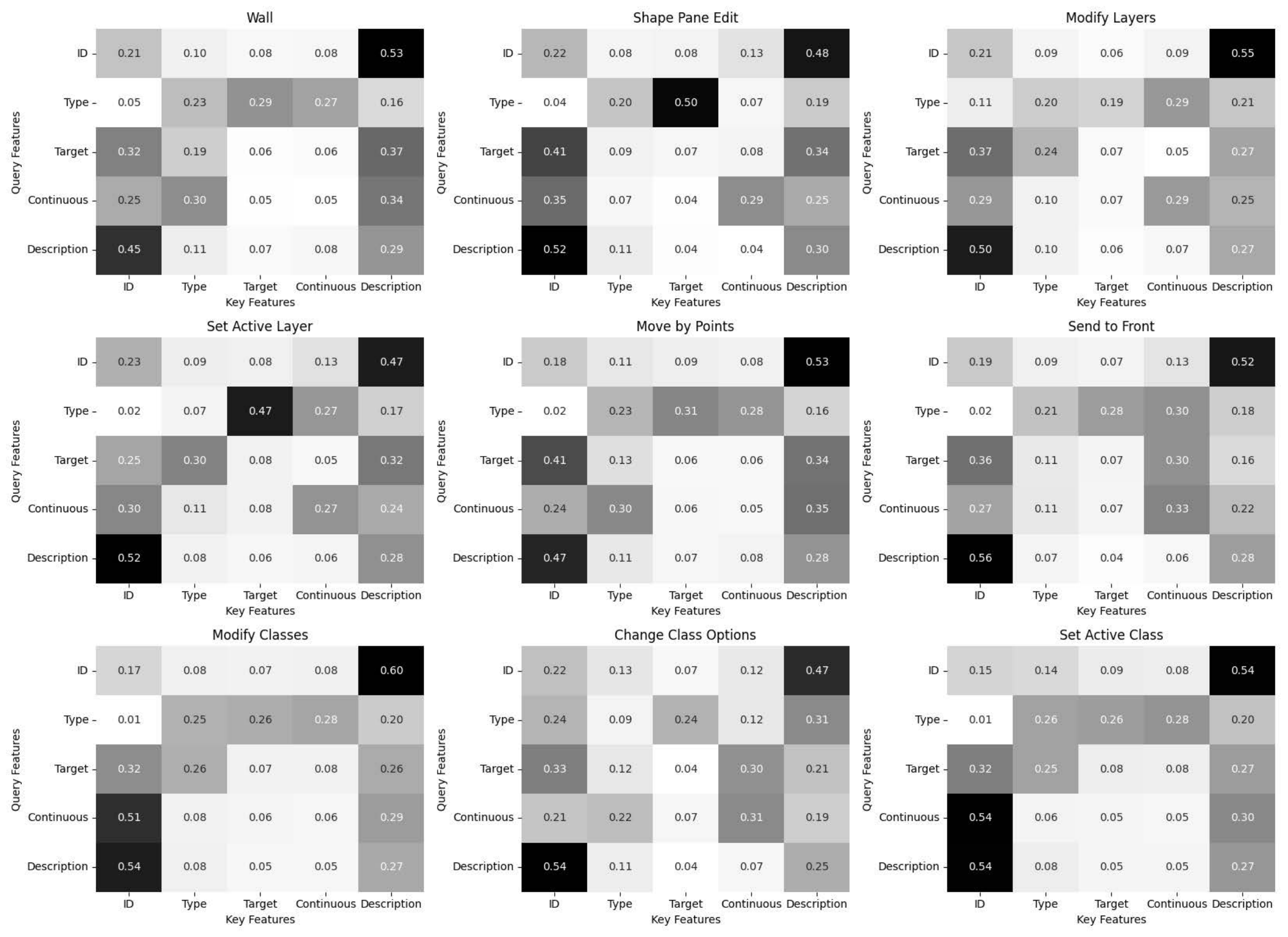}
    \caption{Visualization of attention weights in the feature fusion module attached to \textbf{BERT} backbone, demonstrating distinct feature aggregation patterns for different commands in the given input sequence $[Wall, Shape\, Pane\, Edit, ..., Set\,Active\, Class]$. \textit{Continuous} represents the continuous features (time intervals + consecutive occurrences).}
    \label{fig:bert_feat_fusion}
\end{figure}

\begin{figure}[ht]
    \centering
    \includegraphics[width=1\linewidth]{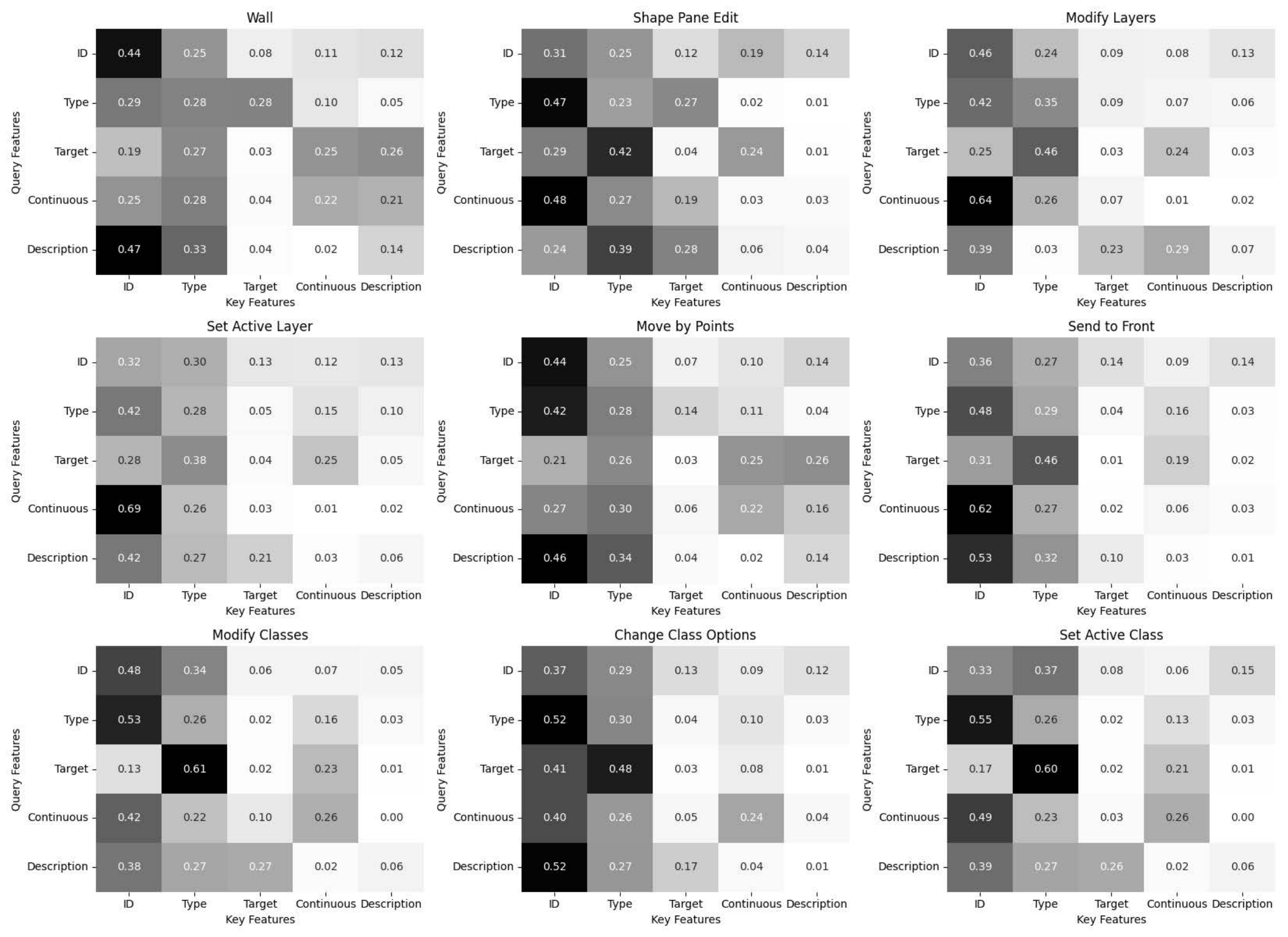}
    \caption{Visualization of attention weights in the feature fusion module attached to \textbf{Mixtral} backbone, demonstrating distinct feature aggregation patterns for different commands in the given input sequence $[Wall, Shape\, Pane\, Edit, ..., Set\,Active\, Class]$. \textit{Continuous} represents the continuous features (time intervals + consecutive occurrences).}
    \label{fig:mixtral_feat_fusion}
\end{figure}

As mentioned in Section \ref{sec:att_fusion}, our model employs attention mechanisms twice: (1) multi-head attention for feature fusion within each command (intra-command), and (2) attention in Transformer blocks for sequence modeling across commands (inter-command). To better understand the learned patterns, we visualized both attention mechanisms and compared different Transformer backbones. Given an input command sequence: 

\textit{[Wall, Shape Pane Edit, Modify Layers, Set Active Layer, Wall, Move by Points, Send to Front, Modify Classes, Change Class Options, Set Active Class]}, 

Figure \ref{fig:bert_feat_fusion} and \ref{fig:mixtral_feat_fusion} illustrate the attention weights assigned to different features when the feature fusion module aggregates information corresponding to each command. Comparing the two figures, an interesting observation is that the feature fusion module prioritizes different aspects when aggregating the five types of command features (ID, Type, Target, Continuous, and Description) depending on the downstream Transformer model. For example, in the case of BERT, the module focuses more on fusing the semantic information from the command description. In contrast, for Mixtral, it places greater emphasis on ID and Type features. For different commands within both figures, the feature fusion module learns unique interaction patterns among their features through the attention mechanism, enabling more effective and dynamic aggregation.

\begin{figure}[ht!]
     \begin{subfigure}[c]{1\linewidth}
        \centering
         \includegraphics[width=1\linewidth]{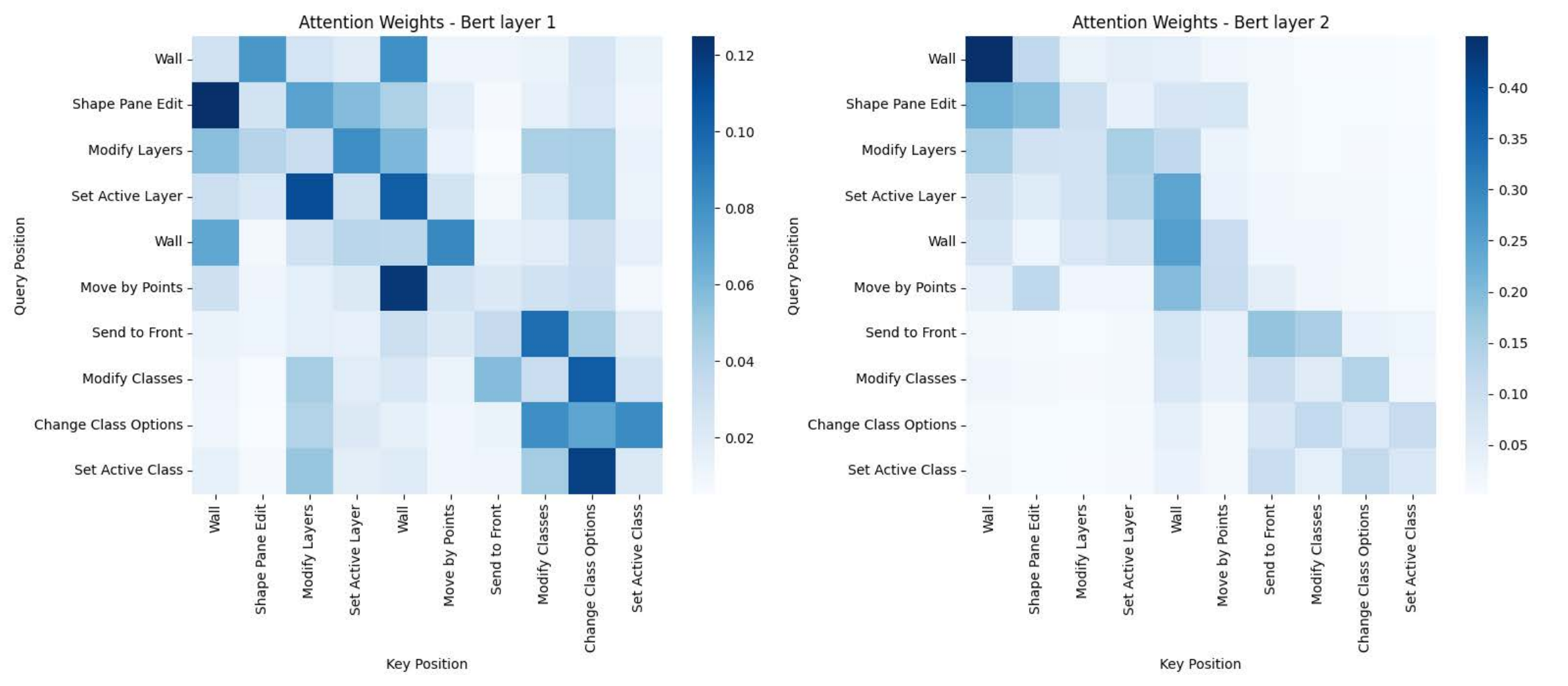}
        \caption{Visualization of attention weights in the \textbf{BERT} layers for the given input command sequence.}
        \label{fig:bert_sub1}
    \end{subfigure}

    \begin{subfigure}[c]{1\linewidth}
        \centering
        \includegraphics[width=1\linewidth]{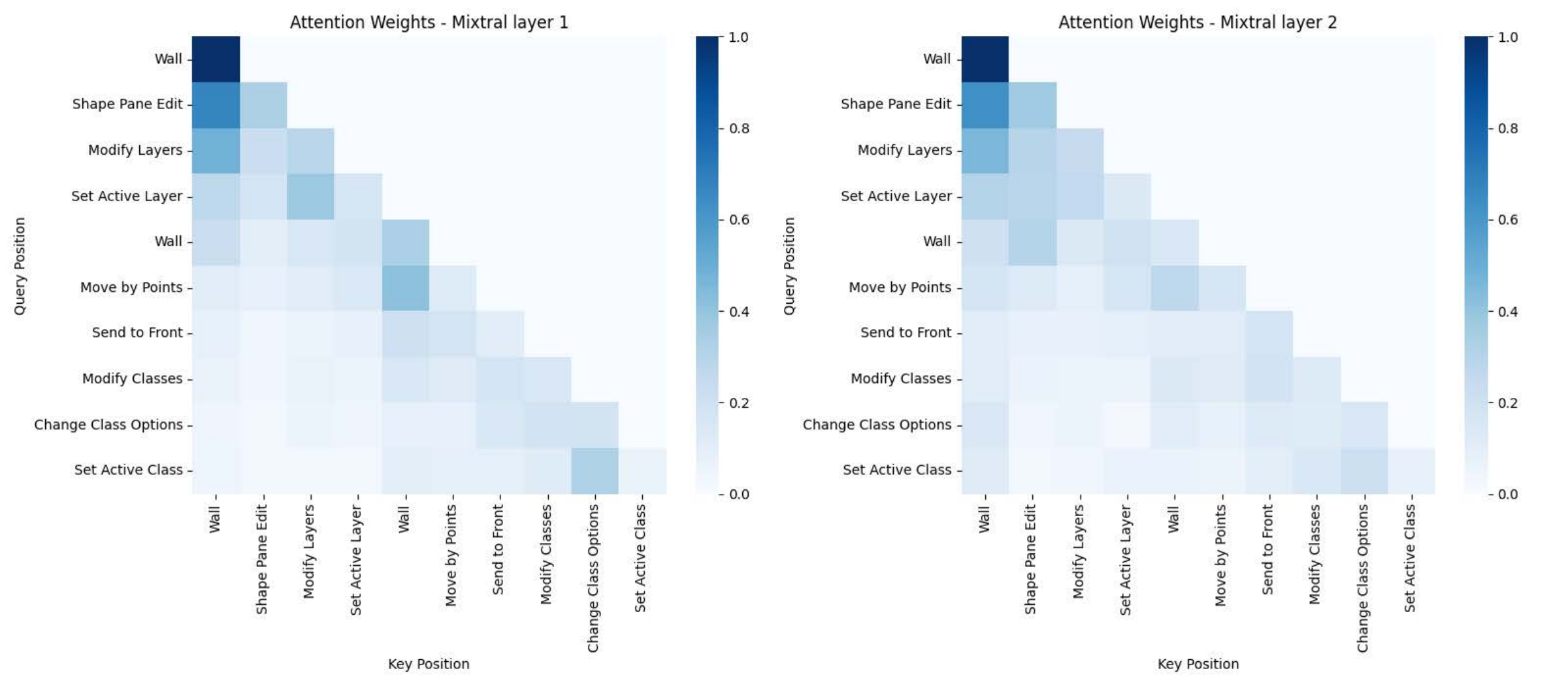}
        \caption{Visualization of attention weights in the \textbf{Mixtral} layers for the given input command sequence.}
        \label{fig:mixntral_sub2}
    \end{subfigure}
    \centering
 \caption{Visualization of attention weights in the two different types of Transformers layers for a given input command sequence $[Wall, Shape\, Pane\, Edit, ..., Set\,Active\, Class]$}
\label{att_layer_fig}
\end{figure}

Figure \ref{att_layer_fig} visualizes the attention patterns of two Transformer models on the given input command sequence. Compared to the unidirectional attention of the Mixtral decoder, the bidirectional attention of the BERT encoder is more dispersed. Both models capture logically relevant command pairs, such as the relationship between \textit{Move by Points} and the preceding \textit{Wall}, as well as the sequential dependency between \textit{Change Class Options} and \textit{Set Active Class}.

\newpage
\section{Deployment and system integration}

\begin{figure}
    \centering
    \includegraphics[width=\linewidth]{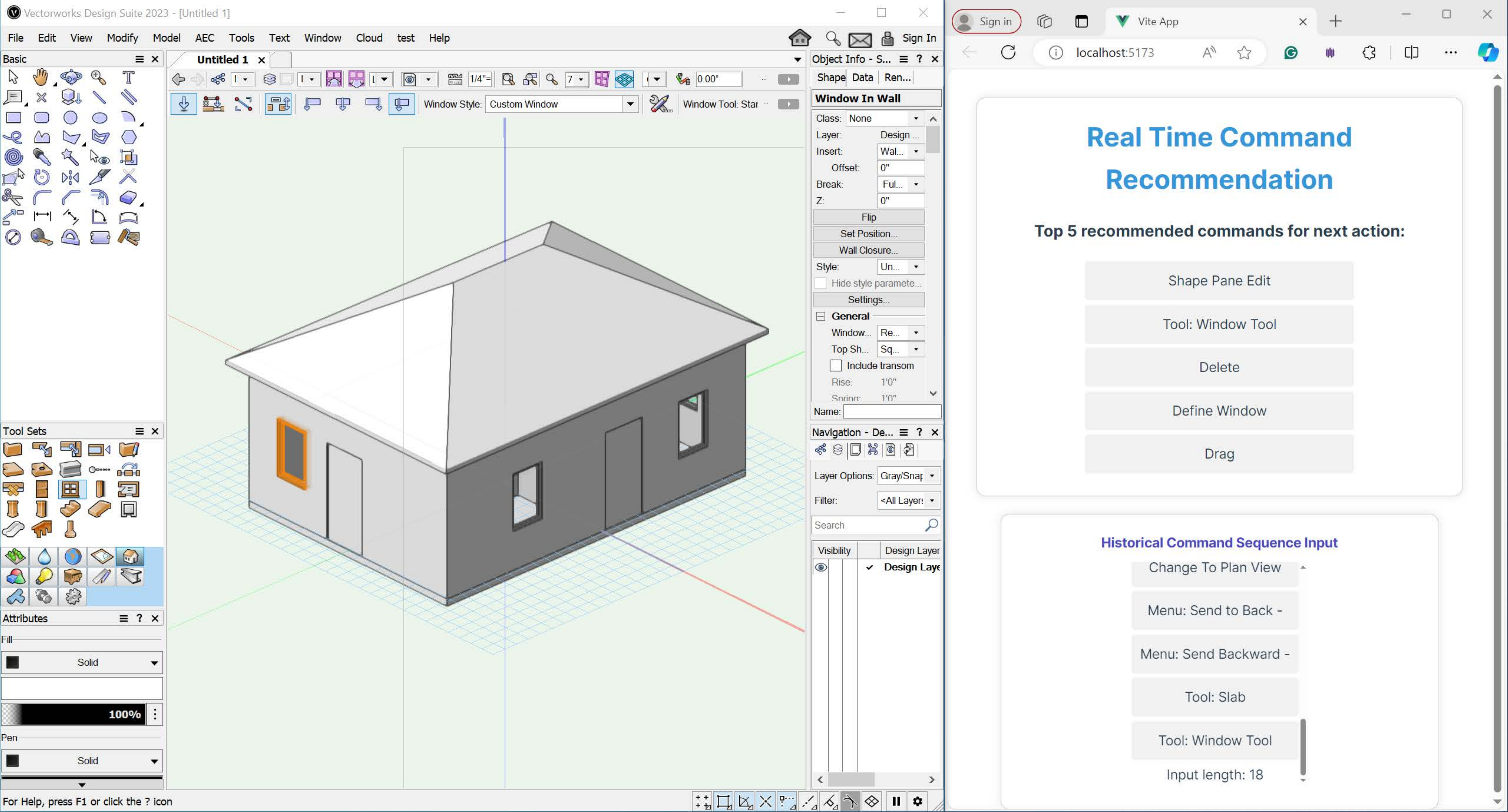}
    \caption{Software prototype running in parallel to Vectorworks, predicting next commands during the BIM authoring process}
    \label{demo}
\end{figure}

\begin{figure}
    \centering
    \includegraphics[width=0.8\linewidth]{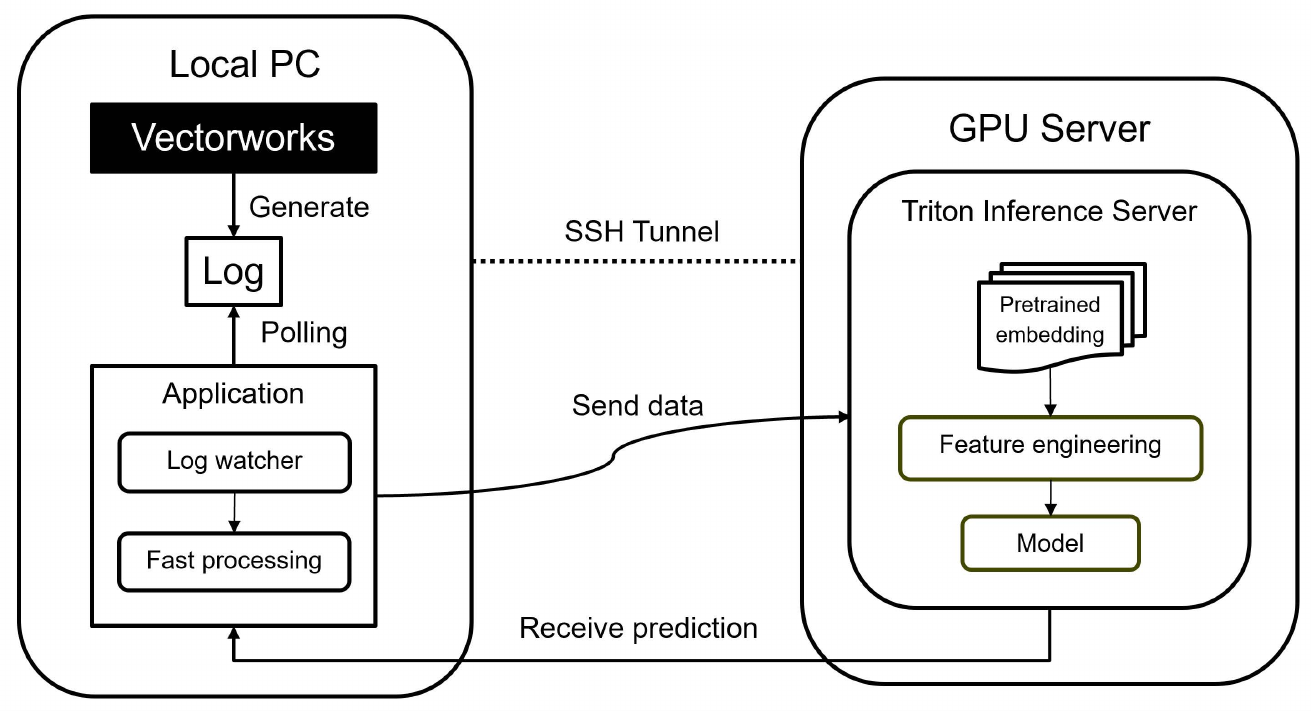}
    \caption{Proposed software architecture of deploying model for real-time command recommendations in Vectorworks. This standalone architecture is not coupled with Vectorworks as a plugin; therefore, it can be extended to other BIM authoring software.}
    \label{software_architecture}
\end{figure}

We developed a software prototype (Figure \ref{demo}) for integrating the proposed method and providing real-time recommendations for the next BIM commands. Figure \ref{software_architecture} illustrates the underlying software architecture.

On a local PC, we implemented an application using Vue.js and FastAPI to monitor and poll real-time log data generated by Vectorworks during the BIM authoring process. In the backend, the pipeline proposed in Section \ref{datafilteringworkflow} was implemented to efficiently process the data. Specifically, we employ the filtering and tracking algorithms introduced in Section \ref{actualmodelingflowtracking} to extract the user's actual operation logic on the fly. Then, we leverage the multilingual dictionary from Section \ref{multilanguagealignment} and the high- and low-level command mapping from Section \ref{redundantcommand} to align command representations in the modeling session and remove redundant entries. Finally, the processed command sequence is sent to a remote server.

Considering the computational cost of model inference, we deploy the trained model and feature engineering pipeline on an Nvidia Triton inference server \cite{NVIDIA_Corporation_Triton_Inference_Server} hosted on a remote GPU server. The GPU server connects to the local PC running Vectorworks via SSH and exchanges data with the local application using the GRPC/HTTP protocol. To mitigate the impact of the large volume of augmented information from the LLM on data transmission speed, we pre-encode command descriptions into semantic embeddings using a pre-trained text embedding model and store them on the GPU server for query by the feature engineering pipeline. Additionally, the feature engineering pipeline encodes categorical information and computes numerical features such as command time intervals and consecutive occurrences, followed by normalization. This pipeline is implemented using the open-source NVTabular library \cite{nvidia_nvtabular}, which supports GPU acceleration and distributed computing, enabling efficient processing of large-scale data. The processed sequence features are fed into the model for inference. The inference results are then transmitted back to our application, where the predicted outcomes are dynamically displayed on the frontend.

\newpage
\section{Discussion and future works}

\subsection{Dataset scale and comparison with previous studies}

The dataset used in this study surpasses the scale of previous research related to design command prediction. Pan et al. \cite{pan2020_log_mining} used BIM logs from an architecture company spanning roughly two years, with a total of 352,056 lines and 14 classes of commands to be predicted. Gao et al. \cite{GAO2022104026} collected BIM logs from a university course on learning 3D modeling software, yielding 490,462 lines with 89 different commands. Matejka et al. from Autodesk \cite{CommunityCommands} collected around 40,000,000 interaction tuples from 16,000 AutoCAD users over six months through the Customer Involvement Program, and used a statistical collaborative filtering algorithm to recommend the next command. In comparison, our final dataset comprises 753,800,816 lines and 4,939 kinds of commands collected in six months. These logs capture anonymous user operations in Vectorworks across different countries, disciplines, and projects, going beyond purely modeling commands to include rendering, lighting, simulation, and more. Such breadth demands more comprehensive and efficient data processing methods and more complex neural network designs. This is also one of our key motivations for adapting modern deep sequential recommendation systems to BIM command prediction.

Traditionally, design behaviors are viewed as highly personalized and complex. However, the promising results on the validation set in our study suggest that designers from different disciplines and regions still exhibit certain shared behavioral patterns, which modern Transformer models can effectively learn from massive data. This finding implies that some sub-processes in BIM authoring might be well-defined and deterministic, serving as essential components of the design process \cite{GAO2022104026}. From another perspective, although designers differ in domain knowledge and discipline, they are essentially using a common “modeling language” when employing BIM authoring software to concretize their design intents. Its vocabulary consists of software commands, and the modern Transformers originally developed for NLP can capture the universal “syntax” of these modeling processes remarkably well.

There might be the concern that inappropriate work sequences may get captured and negatively influence the system. Although individual users, even experienced ones, may sometimes perform suboptimal workflows, it is highly improbable that inappropriate or flawed workflows would be systematically repeated by such a large and diverse user base. Instead, effective and common command sequences will emerge as strong signals, while bad or erroneous sequences will fall into statistical noise and are unlikely to be learned by the model as general patterns.

\subsection{Evaluation strategy}
\label{evaluation_strategy}
In this study, we apply Transformer-based sequential recommendation models to the BIM command prediction task and adopt a self-supervised learning strategy by masking the input sequence to generate prediction targets. This process ensures that at each time step, there is only one ground truth --- the masked target command. Although the model predicts a ranked list of multiple candidate commands, evaluation metrics such as Recall and NDCG primarily focus on the ranking position of the single ground truth within the recommendation list. Such an offline evaluation method is widely used in the field of sequential recommendation \cite{BOKA2024102427,10.1145/3357384.3357895,8594844}. However, it is important to note that while other commands in the list may hold potential value for users in real-world scenarios, the constructed offline data contain only one explicit interaction feedback per instance. Consequently, treating all non-target commands within the list as "negative samples" is a necessary and common simplification/assumption. 

In future work, online evaluation conducted by real-world users would be meaningful and complementary, as multiple correct choices may exist for the next command in real-world design. For instance, further A/B testing or user feedback can help validate the actual utility of other recommended commands, providing a more comprehensive assessment of the model's overall performance.

\subsection{Minority commands}

\begin{figure}[ht!]
    \centering
    \includegraphics[width=0.7\linewidth]{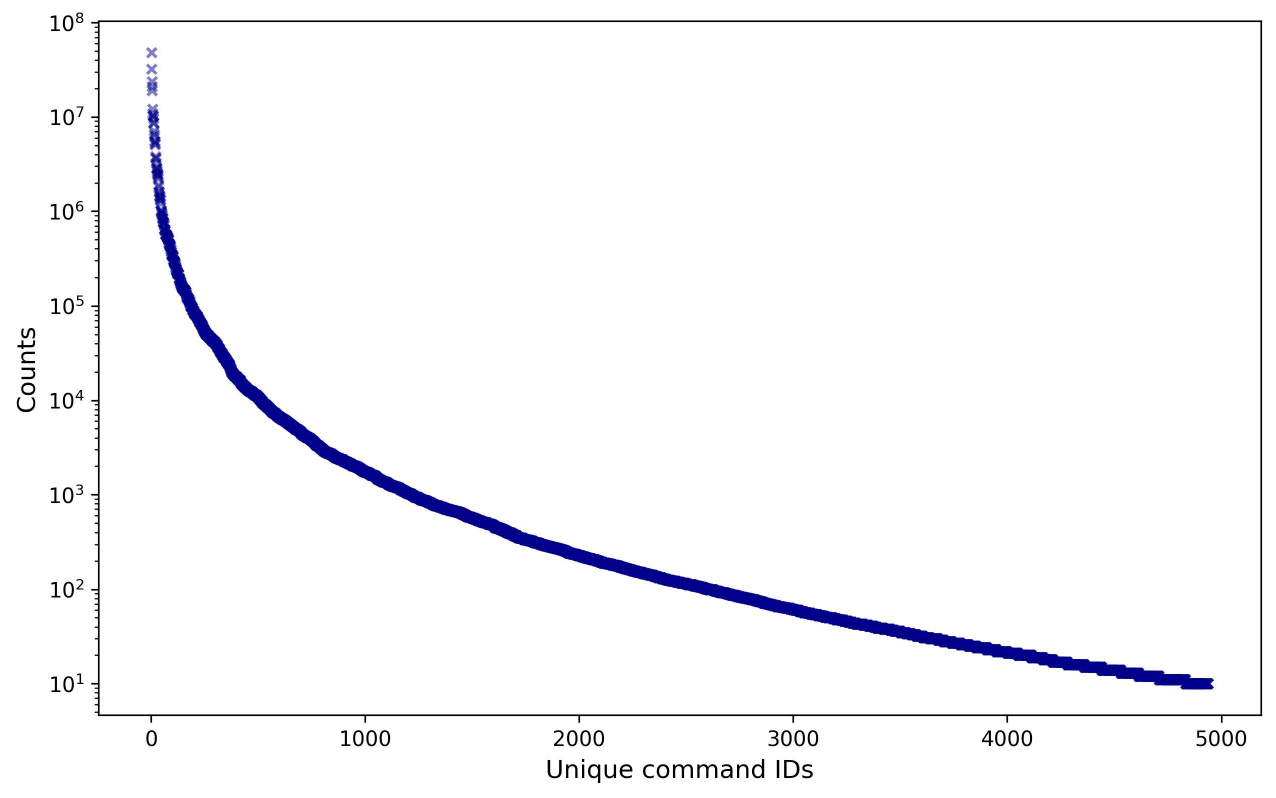}
    \caption{The count distribution of unique command IDs in the final dataset}
    \label{commandsdistributions}
\end{figure}

Previous studies, possibly due to data scale limitations, have largely avoided discussing the severe class imbalance in command distributions within BIM logs. In reality, certain modeling commands, such as \textit{Resize}, \textit{Move}, and \textit{Rotate}, are used with high frequency during the design process. Their occurrence in command sequences is significantly higher than that of niche, domain-specific commands, leading to a long-tail distribution, as illustrated in Figure \ref{commandsdistributions}. This issue becomes particularly pronounced in large-scale BIM log datasets, as the most popular commands and niche commands may differ by several orders of magnitude.

As a result, during the training process, models tend to oversample these mainstream commands while failing to adequately learn the usage patterns of minority commands, leading to suboptimal performance on the validation set. Although we employed focal loss to encourage the model to focus more on hard-to-classify or frequently misclassified samples to mitigate the class imbalance issue, our ablation study indicates that this approach has limited effectiveness.

Future work should focus on optimizing the recommendation of minority, domain-specific commands. One potential strategy is to use the model trained in this study as a \textbf{pretrained base model}, followed by fine-tuning on domain-specific log data (e.g., from structural engineering, landscape design, etc.). Alternatively, adjustments can be made at the data loader level by oversampling minority commands during training or generating synthetic sessions considering long-tail distribution \cite{10.1145/3539618.3591718}, ensuring the model encounters them more frequently during the learning process.

\subsection{Scaling to the production environment}

The current prototype's software architecture is designed for a single local PC. To extend to large-scale multi-user scenarios, communication between local PCs and the inference server requires more complex mechanisms, such as asynchronous messaging or additional routing. Meanwhile, on the server side, the current model deployment strategy and hardware should be adjusted to enable parallel computation across multiple GPU nodes, balancing the large inference workload.

Additionally, the prototype directly polls real-time logs generated by Vectorworks without anonymization. Future deployments must incorporate appropriate methods to protect user privacy. Lastly, software versions undergo periodic iterations, during which new commands are introduced and old ones are deprecated, potentially affecting recommendation accuracy. To address this, the existing approach needs to be extended into a CI/CD pipeline to enable continuous training, evaluation, and deployment of the command recommendation model.

\section{Conclusion}

In this study, we proposed an end-to-end BIM command recommendation framework and tailored a Transformer-based model on a large-scale real-world BIM log dataset collected by Vectorworks. Experimental results demonstrate that the model can learn universal and generalizable modeling patterns from anonymous user interaction sequences across different countries, disciplines, and projects. The key contributions of this work are summarized as follows:
\begin{enumerate}
    \item A comprehensive data filtering and enhancement method is proposed to address real-world engineering challenges in large-scale raw BIM logs, including multilingual content, excessive redundancy, misaligned command IDs, etc. By leveraging LLMs combined with domain knowledge, the approach augments command information in raw logs without relying on custom-developed loggers, thereby generating high-quality training data. A modified BPE algorithm is utilized to generate multi-command workflows, enabling subsequent model to recommend multiple consecutive command steps in a single inference.
    \item A command recommendation model is proposed, incorporating the attention-based feature fusion module, Transformer backbone from the state-of-the-art LLMs, multi-task learning strategy, and focal loss for mitigating data imbalance. The proposed model outperforms baseline approaches across various metrics.
    \item Comprehensive experiments systematically compare different Transformer architectures and conduct an in-depth analysis of the individual model components.
    \item A software prototype is implemented to integrate the trained model into the BIM authoring process, enabling real-time next command recommendation.
\end{enumerate}

 The conceptualization, aesthetics, and functionality of architecture involve highly personalized and complex design decisions.  The proposed system does not attempt to automate such high-level design decisions, or replace the human designers. We emphasize that the quality of the final building design still depends on the designer's knowledge and decisions, while the proposed system supports the software modeling process needed to realize those design intents. Our approach aims to improve user efficiency, reduce modeling time, and lower the learning curve. 

\section{Data availability statement}

The code is available at: \url{https://github.com/dcy0577/BIM-Command-Recommendation.git}. The dataset used in this study is confidential.

\section{Acknowledgment}
This work is funded by the Nemetschek Group, which is gratefully acknowledged. We sincerely appreciate the data and licensing support provided by Vectorworks, Inc.

\newpage
\appendix
\section{Detailed information of the dataset used in this study}
\label{app1}

{\footnotesize
\begin{longtable}{p{2.5cm} p{3cm} c p{5cm}}
\caption{Classification of commands based on pre-defined categories and prefixes in the raw log dataset.}
\footnotesize
\label{table_categories} \\

\toprule

\textbf{Category}  & \textbf{Prefixes} & \textbf{Total number} & \textbf{Explanation} \\
\midrule
\endfirsthead

\toprule
\textbf{Category}  & \textbf{Prefixes} & \textbf{Total number} & \textbf{Explanation} \\
\midrule
\endhead

Tool & Tool & 1,196,303,595 & Refers to an action or process initiated by a tool button, such as creating or editing an object. \\
\midrule
Menu & Menu & 171,695,712 & Indicates an action that was triggered from a menu button, related to a user command from the toolbar or drop-down menu. \\
\midrule
UNDO & Event & 11,522,150,636 & Marks the start of an event.\\
\midrule
UNDO & End Event & 7,499,931,684 & Marks the completion of an event. \\
\midrule
UNDO & DestroyEvent & 7,527,759,023 & Indicates the removal of an event from the undo-able list that reached its maximum length.\\
\midrule
UNDO & Redo Event & 316,153 & Refers to an action where a previously undone operation is redone.\\
\midrule
UNDO & Undo Event & 35,945,199 & Refers to the action of reversing a previous operation or change. \\
\midrule
UNDO & Abort Event & 6,836,454 & Refers to an event that was canceled before completion. \\
\midrule
UNDO & Begin Internal Event & 472,685,421 & Marks the start of an internal process within the software, often related to background tasks or functions. \\
\midrule
UNDO & Beta ForEach Alert & 11 & Refers to a beta testing phase alert related to iterative processes applied to multiple elements. \\
\midrule
UNDO & Beta Undo Alert & 673,901 & Refers to an alert or message generated during a beta phase, specifically related to undo actions. \\
\midrule
UNDO & Project Sharing Problem & 438,244 & Logs an issue or conflict that occurred during the project-sharing process in collaborative workflows. \\
\midrule
UNDO & Undo Problem & 392,054 & Indicates a problem or error encountered during the undo process.\\
\midrule
UNDO & Undo and Remove Action & 3,939,856,492 & Record the undo operation of temporary events, such as the automatic undo operation of quick events like zoom. \\

\bottomrule
\end{longtable}
}

{\scriptsize
\begin{longtable}{cccp{6.5cm}}
\caption{Example commands/workflows with the LLM-augmented meta-information in the final dataset}
\label{merged_table} \\
\scriptsize \\
\toprule
\textbf{Command/Workflow Name}  & \textbf{Types} & \textbf{Target} & \textbf{Description} \\
\midrule
\endfirsthead

\toprule
\textbf{Command/Workflow Name}  & \textbf{Types} & \textbf{Target} & \textbf{Description} \\
\midrule
\endhead

\textbf{Send to Front} 
& Send & Object 
& The entry indicates the execution of the "Send to Front" command, which alters the stacking order of selected objects within the design layer of Vectorworks. This command moves the chosen object to the forefront of the stack, ensuring it appears above any overlapping objects in the drawing. The stacking order is crucial for visual clarity and organization in design presentations. \\
\midrule
\textbf{Wall End Cap} 
& Create & EndCap 
& The Wall End Cap tool in Vectorworks facilitates the creation of both standard and custom end caps for walls. It operates in three distinct modes: Component Wrap, which automatically wraps a selected component; Add, which allows for the addition of a custom shape; and Clip, which enables the removal of a shape from a component to form a custom end cap. This tool ensures that the end caps move with the wall and adjust accordingly when the wall is resized, maintaining their dimensional integrity within wall schedules and exports. \\
\midrule
\textbf{Eyedropper Tool} 
& Copy & Object 
& The log indicates the use of the Eyedropper Tool, which is a feature within the Vectorworks tool palette. This tool allows users to sample and apply attributes from one object to another, facilitating efficient design workflows by streamlining the process of copying properties such as color, texture, and other settings between elements in a project. \\
\midrule
\textbf{Texture} 
& Apply & Texture 
& The log indicates the use of the Texture tool within Vectorworks, which allows users to apply textures to entire objects or specific faces with a simple click. It highlights various modes of operation, such as applying textures to objects or faces, replacing existing textures, and picking up textures from selected objects, all of which can be managed through the Object Info palette's Render tab. The Texture tool is versatile, supporting a range of object types, including generic solids, 3D primitives, and extrusions. Users can also apply textures from the Resource Manager, enhancing workflow efficiency by allowing for direct texture application and management. \\
\midrule
\textbf{Constrained Lin Dim} 
& Create & Dimension 
& The log indicates the use of the Constrained Linear Dimension tool, which allows users to create a dimension line with a single measurement. It outlines the steps for setting measurement points and adjusting the dimension line's orientation in both 2D and 3D views, ensuring that the dimension is constrained to specific axes or aligned with 3D object faces. This functionality is essential for accurately representing dimensions in design projects. \\
\midrule
\textbf{Rectangle; Extrude and Edit} 
& Workflow & Object 
& The log indicates that the user has utilized the Rectangle tool within Vectorworks, specifically accessing the Extrude and Edit menu to convert a 2D rectangle into a 3D form. This operation allows for the specification of height, enabling the user to create complex geometries from simple shapes while leveraging the Push/Pull mode for immediate extrusion in 3D views. Overall, this workflow enhances the modeling capabilities of the software by facilitating the transformation of basic 2D elements into detailed 3D objects. \\
\midrule
\textbf{Copy; Set Active Layer; Paste} 
& Workflow & Object 
& The command captures a sequence of actions within the Vectorworks environment, specifically the execution of the "Copy" command, the setting of the active layer, and the "Paste" command. These operations are crucial for effective design management, allowing users to duplicate elements, specify the layer for modifications, and transfer objects while maintaining class integrity and visibility settings. This workflow exemplifies the fundamental interactions that facilitate efficient project development in Vectorworks. \\
\midrule
\textbf{Rectangle; Add Surface} 
& Workflow & Object 
& The log indicates the use of the Rectangle tool within the Vectorworks environment, specifically highlighting the "Add Surface" command. This operation combines multiple co-planar rectangles into a single entity, provided they are not symbols, are touching or overlapping, and are not locked or grouped. The resulting object inherits properties from the bottom object in the selection stack, streamlining the modeling process for creating polygons. \\
\midrule
\textbf{Line; Move by Points} 
& Workflow & Line 
& The command indicates the utilization of the "Line" tool and the "Move by Points" tool within Vectorworks. The "Line" tool is employed to create linear representations of building materials along a defined path, allowing for customization of attributes such as fill style and line thickness. Meanwhile, the "Move by Points" tool facilitates the movement, duplication, and distribution of selected objects by clicking on specified points, enhancing precision and flexibility in object manipulation during the design process. \\
\bottomrule
\end{longtable}
}

\newpage
\section{Data processing algorithms}
\label{app:alg}

\begin{algorithm}[ht]
\caption{Matching and Removal of Redo and Undo Commands}
\label{algorithm_undo_redo}
\begin{algorithmic}[1]
\State Initialize \(recent\_commands \gets \emptyset\)
\State Initialize \(recent\_undone\_commands \gets \emptyset\)
\State Initialize \(to\_remove\_indices \gets \emptyset\)

\For{each \((\text{cmd}, idx)\) in each log session}
  \If{\(\text{cmd}\) is a normal command}
    \State Append \((\text{cmd}, idx)\) to \(recent\_commands\)
  \ElsIf{\(\text{cmd} = \text{``Undo''}\)}
    \State Search backward in \(recent\_commands\) for a matching command \((c, i)\)
    \If{found}
      \State Add \(idx\) and \(i\) to \(to\_remove\_indices\)
      \State Remove \((c, i)\) from \(recent\_commands\)
      \State Append \((c, i)\) to \(recent\_undone\_commands\)
    \EndIf
  \ElsIf{\(\text{cmd} = \text{``Redo''}\)}
    \State Search backward in \(recent\_undone\_commands\) for a matching command \((c, i)\)
    \If{found}
      \State Remove \(i\) from \(to\_remove\_indices\)
      \State Add \(idx\) to \(to\_remove\_indices\)
      \State Remove \((c, i)\) from \(recent\_undone\_commands\)
      \State Append \((c, i)\) to \(recent\_commands\)
    \Else
      \State Add \(idx\) to \(to\_remove\_indices\)
    \EndIf
  \EndIf
\EndFor

\State Remove all entries with indices in \(to\_remove\_indices\) from the log

\end{algorithmic}
\end{algorithm}

\newpage
\begin{algorithm}[ht]
\caption{Multi-language Alignment}
\label{algorithm_language}
\begin{algorithmic}[1]

\Require Actual modeling logs $\textbf{S}$
\Require Predefined similarity threshold $\phi$
\Ensure Aligned logs $\textbf{S}_{\textit{aligned}}$

\State $\textbf{S}_{\textit{unique}} \gets \textit{dropDuplicates}(\textbf{S})$ \Comment{Filter unique log entries}
\State $\textbf{S}_{\textit{id\_groups}} \gets \textit{groupByCommandID}(\textbf{S}_{\textit{unique}})$

\For{$\textbf{S}_{\textit{id}} \in \textbf{S}_{\textit{id\_groups}}$}
    \For{$\text{s} \in \textbf{S}_{\textit{id}}$}
        \State $\text{s}_{\textit{eng}} \gets \textit{translateToEnglish}(\text{s}_{\textit{message}})$
        \State $\text{s}_{\textit{vector}} \gets \textit{embedding}(\text{s}_{\textit{eng}})$
    \EndFor
    \State $\phi_{\textit{id}} \gets \text{similarityCalculation}(\textbf{S}_{\textit{id}})$
    \If{$\phi_{\textit{id}} \geq \phi$}
        \State $\text{s}_{\textit{centroid}} \gets \frac{1}{|\textbf{S}_{\textit{id}}|} \sum_{\text{s}_{\textit{k}} \in \textbf{S}_{\textit{id}}} \text{s}_{\textit{k}}$ \Comment{Calculate group centroid}
        \State $\textbf{S}_{\textit{id\_unified}} \gets \text{s}_{\textit{centroid\_eng}}$ \Comment{Assign centroid's name as unified command name}
    \Else
        \State $\textbf{S}_{\textit{sub\_id\_groups}} \gets \text{DBSCAN}(\textbf{S}_{\textit{id}})$ \Comment{Sub-clustering}
        \For{$\textbf{S}_{\textit{sub\_id}} \in \textbf{S}_{\textit{sub\_id\_groups}}$}
            \State $\text{s}_{\textit{centroid}} \gets \frac{1}{|\textbf{S}_{\textit{sub\_id}}|} \sum_{\text{s}_{\textit{k}} \in \textbf{S}_{\textit{sub\_id}}} \text{s}_{\textit{k}}$
            \State $\textbf{S}_{\textit{sub\_id\_unified}} \gets \text{s}_{\textit{centroid\_eng}}$ \Comment{Assign centroid's  name as unified command name}
        \EndFor
    \EndIf
\EndFor

\State $\textbf{S}_{\textit{dictionary}} \gets \textbf{S}_{\textit{id}} \cup \textbf{S}_{\textit{sub\_id}}$
\State $\textbf{S}_{\textit{aligned}} \gets \text{languageAlignment}(\textbf{S}_{\textit{dictionary}}, \textbf{S})$
\State \Return $\textbf{S}_{\textit{aligned}}$

\end{algorithmic}
\end{algorithm}

\newpage
\begin{algorithm}[ht]
\caption{Redundant Command Identification}
\label{mapping}
\begin{algorithmic}[1]

\Require Aligned logs \(\textbf{S}_{\textit{aligned}}\)
\Require Similarity threshold \(\phi\)
\Ensure High-/low-level command mapping \(\textbf{S}_{\textit{mapping}}\)

\State \(\textbf{S}_{\textit{mapping}} \gets \emptyset\)
\State \(\textbf{S}_{\textit{high\_level}} \gets \textit{extractHighLevelCommands}(\textbf{S}_{\textit{translated}})\)

\For{\textbf{each} \(\text{s}_{\textit{high\_level}} \in \textbf{S}_{\textit{high\_level}}\)}
    \State \(\textbf{S}_{\textit{pairs}} \gets \emptyset\)
    \State \(\textbf{S}_{\textit{low\_level}} \gets \textit{findFollowingLowLevelCommands}(\text{s}_{\textit{high\_level}}, \textbf{S}_{\textit{aligned}})\)
    \State \(\text{support}_{\textit{high\_level}} \gets \textit{ARMCalculateSupport}(\text{s}_{\textit{high\_level}}, \textbf{S}_{\textit{aligned}})\)

    \For{\textbf{each} \(\text{s}_{\textit{low\_level}} \in \textbf{S}_{\textit{low\_level}}\)}    
        \State \(\text{support}_{\textit{pair}} \gets \textit{ARMCalculateSupport}((\text{s}_{\textit{high\_level}}, \text{s}_{\textit{low\_level}}), \textbf{S}_{\textit{aligned}})\)
        \State \(\text{confidence} \gets \textit{ARMCalculateConfidence}(\text{support}_{\textit{pair}}, \text{support}_{\textit{high\_level}})\)

        \If{\(\text{confidence} > \phi\)}
            \State \(\text{approved} \gets \textit{manualCheck}(\text{s}_{\textit{high\_level}}, \text{s}_{\textit{low\_level}})\)
            \If{\(\text{approved}\)}
                \State \(\textbf{S}_{\textit{pairs}} \gets \textbf{S}_{\textit{pairs}} \cup \{(\text{s}_{\textit{high\_level}}, \text{s}_{\textit{low\_level}})\}\)
            \EndIf
        \EndIf
    \EndFor
    
    \If{\(\textbf{S}_{\textit{pairs}} \neq \emptyset\)}
        \State \(\textbf{S}_{\textit{mapping}} \gets \textbf{S}_{\textit{mapping}} \cup \{\textbf{S}_{\textit{pairs}}\}\)
    \EndIf
\EndFor

\State \Return \(\textbf{S}_{\textit{mapping}}\)

\end{algorithmic}
\end{algorithm}

\newpage
\section{Hyperparameter settings in data filtering and enhancement}
\label{app2}

In this section, we explain the rationale behind our settings of two hyperparameter values in the data filtering and enhancement pipeline: the similarity threshold for multi-language alignment, as described in Section \ref{multilanguagealignment}, and the confidence threshold for association rule mining (ARM), as discussed in Section \ref{redundantcommand}.

In multi-language alignment, if the similarity threshold is too low, we may miss cases where the same ID corresponds to semantically different commands. On the other hand, if the threshold is too high, unsupervised clustering may incorrectly separate semantically similar commands into different clusters due to slight variations in translation. For instance, in ID 424, the translated versions contain five instances of \textit{End Event: Edit Subdivision (424)} and one instance of \textit{End Event: Subdivision Change (424)}. However, the command \textit{End Event: Subdivision Change (424)} may be clustered into a different group, despite referring to the same semantic action.

\begin{figure}[ht!]
    \centering
    \includegraphics[width=1.0\linewidth]{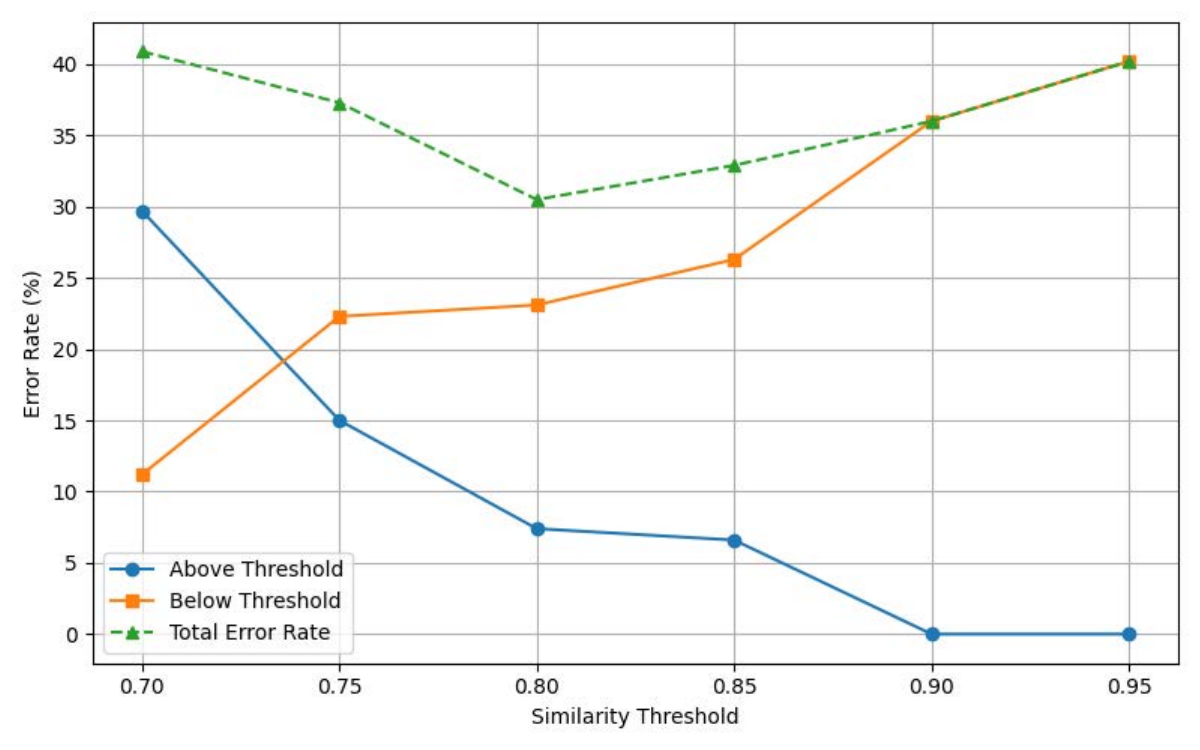}
    \caption{Sensitivity analysis of similarity threshold}
    \label{fig:threhold}
\end{figure}

We tested five different similarity thresholds: 0.95, 0.90, 0.85, 0.80, and 0.75. For each threshold, command IDs with similarity above the threshold were directly labeled, while those below the threshold were processed using unsupervised clustering. To evaluate the quality of this process, we performed stratified sampling \cite{neyman1992two} by selecting 10\% of the IDs from both the above-threshold and below-threshold groups, due to the large size of the dataset. We then manually inspected the sampled results and calculated the error rate. Errors were defined as either (1) incorrect clustering of semantically different commands or (2) misclustering of semantically similar commands. The result can be shown in Figure \ref{fig:threhold}. 

The error rate is lowest around the threshold of 0.80. In this study, we applied a slight buffer and set the threshold to 0.82. However, the overall error rate remained relatively high, at approximately 30\%. To further enhance accuracy, we conducted an additional manual verification step. This involved reviewing and refining the command dictionary by checking whether the translations were accurate and semantically aligned.

The confidence in ARM measures the strength of the relationship between the two commands in a command pair. A higher confidence value indicates a stronger association: when the first command occurs, the corresponding command in the pair is highly likely to be triggered afterward. For example, the confidence of the command pair \textit{Tool: Line (-201)} and \textit{End Event: Create Line (166)} is 0.93, indicating a strong association between them.

However, due to the large size of the dataset, a single command can form a large number of command pairs. Additionally, some commands do not trigger any follow-up commands. For example, \textit{Menu: Close (-4)} forms around 40,000 command pairs after ARM, yet none of them have a confidence value greater than 0.4. We further tested this command in the software environment and confirmed that it does not trigger any follow-up commands. Thus, to reduce the workload caused by the abundance of obviously irrelevant command pairs, we applied a confidence threshold to automatically filter them out. Specifically, we set a relatively low threshold of 0.4, which effectively eliminates the most apparent non-triggering relationships and significantly reduces the need for manual verification. After this filtering step, we further conducted manual refinement within the software to validate and improve the remaining high-confidence command pairs.

Overall, the settings of these two hyperparameters have minimal impact on the final processed data, as all results are subject to additional manual review.

\newpage
\section{Statistical significance test results between Proposed and Baseline models}
\label{app:statis}

\begin{table}[ht!]
\centering

\caption{Statistical significance test results between Proposed and Baseline models using paired bootstrap [\%]}
\label{tab:significance_test}
\renewcommand{\arraystretch}{1.5}
\adjustbox{width=\textwidth}{%
\begin{tabular}{|l|c|c|c|c|c|c|}
\hline
\multirow{2}{*}{\textbf{Backbone}} & \multicolumn{3}{c|}{\textbf{Recall@k}} & \multicolumn{3}{c|}{\textbf{NDCG@k}} \\
\cline{2-7}
& \textbf{k=3} & \textbf{k=5} & \textbf{k=10} & \textbf{k=3} & \textbf{k=5} & \textbf{k=10} \\
\hline
\hline
\multicolumn{7}{|l|}{\textit{\textbf{BERT (Encoder-only)}}} \\
\hline
Mean Difference & 0.3714 & 0.3043 & 0.1026 & 0.4119 & 0.3881 & 0.3225 \\
95\% CI & [0.3191, 0.4244] & [0.2577, 0.3520] & [0.0633, 0.1413] & [0.3728, 0.4506] & [0.3551, 0.4206] & [0.2944, 0.3509] \\
\textit{p}-value (two-sided) & $<0.0001$ & $<0.0001$ & $<0.0001$ & $<0.0001$ & $<0.0001$ & $<0.0001$ \\
\hline
\hline
\multicolumn{7}{|l|}{\textit{\textbf{Mixtral-MoE (Decoder-only)}}} \\
\hline
Mean Difference & 1.2249 & 0.9716 & 0.5847 & 1.2646 & 1.1610 & 1.0342 \\
95\% CI & [1.1659, 1.2825] & [0.9204, 1.0220] & [0.5436, 0.6253] & [1.2210, 1.3078] & [1.1239, 1.1966] & [1.0029, 1.0660] \\
\textit{p}-value (two-sided) & $<0.0001$ & $<0.0001$ & $<0.0001$ & $<0.0001$ & $<0.0001$ & $<0.0001$ \\
\hline
\hline
\multicolumn{7}{|l|}{\textit{\textbf{Llama3.2 (Decoder-only)}}} \\
\hline
Mean Difference & 1.2533 & 0.8964 & 0.6013 & 1.2226 & 1.0758 & 0.9800 \\
95\% CI & [1.1944, 1.3118] & [0.8445, 0.9489] & [0.5597, 0.6425] & [1.1782, 1.2675] & [1.0377, 1.1137] & [0.9473, 1.0126] \\
\textit{p}-value (two-sided) & $<0.0001$ & $<0.0001$ & $<0.0001$ & $<0.0001$ & $<0.0001$ & $<0.0001$ \\
\hline
\hline
\multicolumn{7}{|l|}{\textit{\textbf{T5 (Encoder-Decoder)}}} \\
\hline
Mean Difference & 0.4802 & 0.3756 & 0.2446 & 0.4920 & 0.4497 & 0.4064 \\
95\% CI & [0.4271, 0.5339] & [0.3292, 0.4220] & [0.2066, 0.2825] & [0.4540, 0.5313] & [0.4164, 0.4821] & [0.3783, 0.4348] \\
\textit{p}-value (two-sided) & $<0.0001$ & $<0.0001$ & $<0.0001$ & $<0.0001$ & $<0.0001$ & $<0.0001$ \\
\hline
\end{tabular}%
}
\begin{flushleft}
\scriptsize
- Mean Difference: Average performance improvement of Proposed over Baseline model (Proposed - Baseline) across 10,000 bootstrap samples \\
- 95\% CI: 95\% confidence interval from 10,000 bootstrap iterations \\
- \textit{p}-value computed via permutation test \cite{10.5555/1196379}
\end{flushleft}
\end{table}

To assess whether the performance differences between our proposed models and the baselines are statistically significant, we employed paired bootstrap to calculate the confidence intervals and paired permutation tests to obtain p-values.

\textbf{Bootstrap confidence intervals}. For each evaluation metric, we: 
(1) Calculated the paired differences between the two models: $\Delta_i = X_{1i} - X_{2i}$, where $X_{1i}$ and $X_{2i}$ represent the metric values for the $i$-th test sample from the proposed and baseline models, respectively. 
(2) Generated 10,000 bootstrap samples by resampling the differences with replacement. 
(3) Computed the mean difference for each bootstrap sample and derived the 95\% confidence interval using the 2.5th and 97.5th percentiles of the bootstrap distribution.

\textbf{Paired permutation test.} To obtain p-values, we conducted paired permutation tests \cite{10.5555/1196379} that account for the paired nature of our comparisons (same test samples evaluated by both models). The procedure is as follows: 
(1) Computed the observed test statistic: $T_{\text{obs}} = \text{mean}(X_1) - \text{mean}(X_2)$ 
(2) Under the null hypothesis of no difference between models, we randomly exchanged the model assignments within each pair with probability 0.5, repeated 10,000 times. 
(3) For each permutation $\pi$, calculated the test statistic: $T_\pi = \text{mean}(X_{1\pi}) - \text{mean}(X_{2\pi})$ (4) Computed the two-sided p-value as: $p = P(|T_\pi| \geq |T_{\text{obs}}|)$

The detailed results in Table \ref{tab:significance_test} demonstrate that the improvements of the proposed models over the baselines are statistically significant.

\newpage
\bibliographystyle{elsarticle-num} 
\bibliography{reference}





\end{document}